\documentclass[onecolumn,oneside,11pt,altaffilletter,tightenlines,showpacs,showkeys,
               notitlepage,preprintnumbers,nofootinbib,superscriptaddress,a4paper,floatfix]{revtex4-2}
\pdfoutput=1
\usepackage{graphicx}
\usepackage[centering,hmargin=2.4cm,vmargin=2.4cm]{geometry} 
\usepackage[T1]{fontenc}
\usepackage{color}
\usepackage[normalem]{ulem}
\usepackage{amsmath,amssymb,slashed}
\usepackage[range-units=single,range-phrase=--]{siunitx}
\usepackage{bbm}
\usepackage{epsfig}
\usepackage{grffile,graphicx}\graphicspath{{figs/}} 

\usepackage{enumerate,enumitem}
\usepackage{multirow}
\usepackage{placeins}
\usepackage[dvipsnames]{xcolor}
\usepackage{epstopdf}
\usepackage[utf8]{inputenc}
\usepackage{tikz}
\usepackage{csquotes}
\usepackage{accents} 
\usepackage{booktabs} 
\usepackage{titlesec} 
\usepackage[subfigure,titles]{tocloft} 
\usepackage[colorlinks=true,citecolor=purple,linkcolor=blue]{hyperref}

\usetikzlibrary{trees}
\usetikzlibrary{decorations.pathmorphing}
\usetikzlibrary{decorations.markings}

\definecolor{jblue}  {RGB}{20,50,100}
\definecolor{npurple}  {RGB} {153, 51, 204}
\definecolor{wred}   {RGB}{217,0,56}
\definecolor{white}   {RGB}{255,255,255}

\definecolor{korange}   {RGB}{235, 80,  43}
\definecolor{korange2}   {RGB}{245, 100,  63}
\definecolor{kyelloworange}   {RGB}{255, 210,  110}
\definecolor{kyelloworange2}   {RGB}{240, 170,  90}
\definecolor{kred}   {RGB}{204,  102, 153}
\definecolor{kpurple}   {RGB}{153,  61, 190}
\definecolor{kpurplelight}   {RGB}{213,  161, 230}

\usepackage{cleveref}
\usepackage{subfigure}
\usepackage{lipsum}
\definecolor{red}{rgb}{1.0, 0, 0}
 \usepackage{gensymb}
\allowdisplaybreaks

\titleformat*{\section}{\centering\bfseries\scshape}
\titlelabel{\thetitle\quad}
\titleformat*{\paragraph}{\bfseries}
\titlespacing*{\paragraph}{0pt}{3.25ex plus 1ex minus .2ex}{1em}

\newcommand{\BR}{\text{BR}}

\renewcommand{\vec}[1]{{\mathbf{#1}}}
\newcommand{\iso}[2]{{\ensuremath{{}^{#2}}\ensuremath{\rm #1}}}

\newcommand{\GeV}{\,\mathrm{GeV}}

\newcommand{\Ap}{A^{\prime}}
\newcommand{\Zp}{Z^{\prime}}
\newcommand{\mAp}{m_{\Ap}}
\newcommand{\mZp}{m_{\Zp}}

\usepackage{array}
\newcolumntype{L}[1]{>{\raggedright\let\newline\\\arraybackslash\hspace{0pt}}m{#1}}
\newcolumntype{C}[1]{>{\centering\let\newline\\\arraybackslash\hspace{0pt}}m{#1}}
\newcolumntype{R}[1]{>{\raggedleft\let\newline\\\arraybackslash\hspace{0pt}}m{#1}}

\newlength{\dhatheight}
\newcommand{\doublehat}[1]{%
    \settoheight{\dhatheight}{\ensuremath{\hat{#1}}}%
    \addtolength{\dhatheight}{-0.35ex}%
    \hat{\vphantom{\rule{1pt}{\dhatheight}}%
    \smash{\hat{#1}}}}

\makeatletter
\renewcommand{\p@subsection}{}
\makeatother


\begin{document}
\title{Searching for Physics Beyond the Standard Model \\
       in an Off-Axis DUNE Near Detector}
\author{Moritz~Breitbach} \email{breitbach@uni-mainz.de}
\affiliation{PRISMA Cluster of Excellence and
             Mainz Institute for Theoretical Physics,
             Johannes Gutenberg-Universit\"{a}t, Staudingerweg 7, Mainz, Germany}

\author{Luca~Buonocore}   \email{lbuono@physik.uzh.ch}
\affiliation{Physik Institut, Universit\"at Z\"urich, Switzerland}

\author{Claudia~Frugiuele}\email{claudia.frugiuele@cern.ch}
\affiliation{INFN, Sezione di Milano, Via Celoria 16, I-20133
Milano, Italy.}

\author{Joachim~Kopp}     \email{jkopp@cern.ch}
\affiliation{Theoretical Physics Department, CERN, Geneva, Switzerland}
\affiliation{PRISMA Cluster of Excellence and
             Mainz Institute for Theoretical Physics,
             Johannes Gutenberg-Universit\"{a}t Mainz, Germany}

\author{Lukas~Mittnacht}  \email{lmittna@uni-mainz.de}
\affiliation{PRISMA Cluster of Excellence and
             Mainz Institute for Theoretical Physics,
             Johannes Gutenberg-Universit\"{a}t Mainz, Germany}

\date{\today}

\preprint{MITP-21-004, ZU-TH-5/21, CERN-TH-2021-018}

\begin{abstract}
  Next generation neutrino oscillation experiments like DUNE and T2HK are
  multi-purpose observatories, with a rich physics program beyond oscillation
  measurements.  A special role is played by their near detector facilities,
  which are particularly well-suited to search for weakly coupled dark sector
  particles produced in the primary target.  In this paper, we demonstrate this
  by estimating the sensitivity of the DUNE near detectors to the scattering of
  sub-GeV DM particles and to the decay of sub-GeV sterile neutrinos
  (``heavy neutral leptons'').  We discuss in particular the importance of
  the DUNE-PRISM design, which allows some of the near detectors to be moved
  away from the beam axis. At such off-axis locations, the signal-to-background
  ratio improves for many new physics searches.  We find that this leads to
  a dramatic boost in the sensitivity to boosted DM particles interacting
  mainly with hadrons, while for boosted DM interacting with leptons,
  data taken on-axis leads to marginally stronger exclusion limits. Searches
  for heavy neutral leptons perform equally well in both configurations.
\end{abstract}

\maketitle

\vspace{-0.5cm}
\newcommand{\contentsname}{}
\setlength{\cftbeforesecskip}{0pt}

{
    \hypersetup{hidelinks}
    \tableofcontents
}

\section{Introduction}
\label{sec:intro}

The near detectors of long-baseline neutrino experiments, once considered an afterthought to reduce systematic uncertainties in oscillation measurements, are nowadays independent experiments in their own right. Besides delivering a wealth of data on neutrino interaction physics, it has been realized that they could also serve to probe the existence of physics beyond the Standard Model (SM) \cite{Batell:2009di} such as heavy neutral leptons or other new particles, possibly connected to the dark matter (DM) puzzle \cite{ deNiverville:2011it, deNiverville:2012ij, Dharmapalan:2012xp, Batell:2014yra, Soper:2014ska, Coloma:2015pih, beam1,beam2,beam3,deNiverville:2018dbu,millicharged,trident,DeRomeri:2019kic,Buonocore:2019esg,Batell:2019nwo,Ballett:2019bgd, MicroBooNE:2019izn, Gorbunov:2020rjx, Arguelles:2021dqn}. Indeed, signatures of new dark particles at these experiments can be linked in a predictive way to compelling scenarios of light DM. For instance, ``invisible'' decays of new sub-GeV particles that mediate light DM--SM interactions, can be searched for at neutrino fixed target facilities by looking for scattering of the decay products off nucleons and/or electrons in the near detector. A light DM program at neutrino facilities could complement the next generation light DM direct detection program \cite{cosmicvision}, in particular, the upcoming experiment SENSEI \cite{sensei}. So far, only the Fermilab-based neutrino experiment MiniBooNE has performed dedicated searches for light DM \cite{miniboone,minibooneE} and very recently MicroBooNE has released the first search for heavy neutral leptons (HNL) \cite{MicroBooNE:2019izn} and for a light Higgs decaying into $e^+e^-$~\cite{MicroBooNE:2021ewq}, but the untapped potential is big, with many past and ongoing experiments having the capability to supersede MiniBooNE's sensitivity \cite{beam1,beam2,deNiverville:2018dbu,Buonocore:2019esg}. Importantly, these searches can typically be done fully parasitically to the main neutrino program \cite{beam2,deNiverville:2018dbu}.

The near detector physics program will be taken to the next
level by the DUNE-PRISM detectors, to be installed \SI{574}{m} downstream
from the target \cite{DUNE:2021tad} at the long-baseline neutrino facility (LBNF) at Fermilab, the neutrino source for the DUNE experiment. These detectors -- a liquid argon time
projection chamber (TPC) and a magnetized gaseous argon TPC -- will be mounted
on a movable platform, allowing them to be displaced up to \SI{30.5}{m}
(\SI{53}{mrad}) away from
the beam axis.  This capability mainly serves the detectors' primary purpose,
namely constraining the unoscillated neutrino flux and measuring the neutrino
cross sections. In particular, the neutrino spectrum changes as a function of
the off-axis angle (for kinematic reasons), while the cross sections obviously
do not. Therefore, taking data at different off-axis positions will allow
DUNE-PRISM to disentangle the uncertainties in the neutrino spectrum from the
uncertainties in the neutrino cross section.

In this paper, we will discuss the impact of the DUNE-PRISM concept on
searches for physics beyond the SM.  More specifically, we will
consider the production of light ($\lesssim \si{GeV}$) and very weakly
interacting new particles in the target, followed by their interaction or
decay inside the DUNE-PRISM detectors.  

Among the numerous extensions of the SM that can be probed in
DUNE-PRISM and other accelerator neutrino experiments, we will consider in
particular: (1) light ($\lesssim \si{GeV}$) DM particles produced with
a large Lorentz boost and detectable via dark photon-mediated DM--electron
scattering;   (2) light DM particles detectable via DM--nucleus
scattering (leptophobic DM); and (3) heavy neutral leptons (sterile neutrinos) decaying to
various combinations of neutrinos, charged leptons, and hadrons. In the
following sections, we will introduce these scenarios one by one and discuss
the anticipated sensitivity of DUNE-PRISM, both on-axis and off-axis. Specifically,
\cref{sec:dark-photon} will be focused on dark photon-mediated DM, \cref{sec:leptophobic} will deal with leptophobic DM,
and \cref{sec:hnl} will be about heavy neutral leptons.  We will
discuss our findings and conclude in \cref{sec:summary}. Let us comment
that similar searches for DUNE have been considered previously in
refs.~\cite{beam2,Ballett:2019bgd,Krasnov:2019kdc,Coloma:2020lgy} for the case of on-axis
detectors, and for additional data taking away from the beam axis in
ref.~\cite{DeRomeri:2019kic}.

We will go beyond these studies in two important ways:
\begin{itemize}
  \item We will investigate the usefulness of taking DUNE off-axis data for
    additional scenarios, for which this has never been done before. In
    particular, we will consider scattering of light, leptophobic DM and decays of heavy neutral leptons.

  \item We will also reconsider DM scattering on electrons,
    previously studied in ref.~\cite{DeRomeri:2019kic}. As a cross-check,
    we will reproduce the total rates analysis carried out in this reference,
    but we will also show that an analysis including the electron recoil
    spectrum is equally sensitive on-axis and off-axis.
\end{itemize}

\section{Light Dark Matter Interacting via a Dark Photon}
\label{sec:dark-photon}

Most direct searches for DM lose sensitivity at DM masses below a few
GeV, motivating a new experimental program that focuses specifically on this mass 
range~\cite{cosmicvision}.

One of the simplest and most generic models for DM in the MeV--GeV mass range
augments the SM by a scalar DM particle $\phi$ (or a Majorana fermion)
and a new $U(1)'$ gauge boson, $A'$.  The relevant terms in the Lagrangian
read
\begin{align}
\mathcal{L}_{ \rm DM}=\mathcal{L}_{A^{\prime}}+\mathcal{L}_\phi \,,
\end{align}
with
\begin{align}
\mathcal{L}_{A^{\prime}} =- \frac{1}{4} F'_{\mu \nu}F^{\prime \mu \nu} +\frac{m^2_{A'} }{2}A^{\prime \mu} A^{\prime}_{ \mu}-\frac{1}{2} \epsilon \,  F^{\prime}_{\mu \nu} F^{\mu \nu}\,,\label{eq:L-dark-photon}
\end{align}
and
\begin{align}
\mathcal{L}_\phi =  i g' A^{\prime \mu} J_{\mu}^{\phi}+ (\partial_{\mu} \phi^\dagger) (\partial^{\mu} \phi) - m_{\phi}^2 \phi^\dagger \phi\,,
\end{align}
where $J_{\mu}^{\phi}=  \left[ (\partial_\mu \phi^\dagger) \phi  -  \phi^\dagger  (\partial_\mu \phi \right)]$ is the DM current, $g'$ is the $U(1)'$ gauge coupling, $F_{\mu\nu}$ and $F_{\mu\nu}'$ are the $U(1)$ and $U(1)'$ field strength tensors, respectively, and $\epsilon$ parameterizes the small kinetic mixing between the dark and visible photons. $\epsilon$ thus ultimately controls the interaction strength between the dark photon and SM particles.
For DM lighter than half the dark photon mass ($m_\phi < m_{A'} / 2$),
the thermal relic abundance of $\phi$ is determined by its annihilation cross section
to SM fermions,
\begin{align}
    \sigma(\phi \phi^\dag \to f \bar f) v_\text{rel}
      \sim \frac{8 \pi v_\text{rel}^2 Y}{m_{\phi}^2} \,,
    \label{eq:sigma-ann}
\end{align}
where we have defined the effective coupling strength
\begin{align}
    Y \equiv \epsilon^2 \alpha_D \bigg(\frac{m_\phi}{m_{A'}}\bigg)^4 \,.
\end{align}
As usual, $v_\text{rel}$ denotes the relative velocity of the two annihilating DM particles,
and $\alpha_D \equiv g'^2 / (4\pi)$. 
In the following, we will present our results in the $m_\phi$--$Y$ plane since
this choice makes it easiest to highlight those regions of parameter space where
the correct DM thermal abundance is obtained \cite{Gordan,cosmicvision}. 

One important feature of the annihilation cross section in \cref{eq:sigma-ann}
is its $v_\text{rel}^2$ suppression, through which strong
constraints based on precise measurements of the temperature anisotropies of the
cosmic microwave background radiation \cite{Lin:2011gj, Ade:2015xua, Slatyer:2015jla, 
Slatyer:2015kla} are avoided.
For models with unsuppressed annihilation, these constraints would
rule out thermal freeze-out production of DM candidates with a mass below
$\sim \SI{10}{GeV}$. Thanks to the velocity suppression, scalars or Majorana fermions
can account for the totality of the DM abundance via thermal freeze-out even at masses
below \SI{1}{GeV}.

The kinetic mixing term in \cref{eq:L-dark-photon} implies that any process
that can create a photon can also
create a dark photon, provided this is kinematically allowed. In a meson
production target like the ones employed in neutrino beam experiments, dark
photons can be copiously produced in meson decays such as $\pi^0 \to \gamma A'$
or $\eta \to \gamma A'$, with a smaller contribution from bremsstrahlung.  The
dark photon couples to the dark current with coupling strength $g'$, and to the
SM electromagnetic current with coupling strength $\epsilon\,e$.  We will
consider in particular the case where the dark photon mass, $m_{A'}$, is larger
than twice the DM mass, $m_\phi$. In this case, any $A'$ produced in
the target will rapidly decay, almost exclusively to $\phi \phi^\dag$.  
Hence, a beam of $\phi$ particles will travel alongside the neutrino beam
and eventually reach the near detector, where $\phi$ particles can scatter
on nuclei and electrons.  It is in particular the latter channel --
$\phi$--electron scattering -- that we will focus on because
in this channel neutrino-induced backgrounds are smaller~\cite{deNiverville:2011it,
deNiverville:2018dbu, Buonocore:2019esg}.

\subsection{Dark Matter Production and Detection}
\label{sec:dp-production}

In a proton beam dump, dark photons with masses below $\sim \SI{1}{GeV}$ are mainly produced in the decays of the lightest neutral mesons, $\pi^0$ and $\eta$, and in proton bremsstrahlung via the process $p p \to p p \Ap$. Production processes induced by leptonic secondary particles and their bremsstrahlung are usually subdominant, even if not completely negligible as reported in a detailed calculation in ref.~\cite{Celentano:2020vtu}. We do not consider the latter type of processes in this work, and therefore our estimates should be considered conservative in this regard. In the mass window considered, production mechanisms that can be described in perturbative quantum chromodynamics (QCD), such as Drell--Yan production, are negligible and are not taken into account here. In the simulation of the dark photon signal, we realistically take into account correlations between the geometric acceptance of the detector and the angular spread of the DM flux. For the detector geometry, we consider for simplicity a cylindrical shape oriented along the beam axis with transverse surface given by a circle of radius $\SI{3.5}{m}$.  While this simplified detector model does not exactly match the envisioned geometry of the DUNE-PRISM detectors, the error introduced by our approximation should be negligible compared to the intrinsic uncertainties of the flux prediction.

\paragraph*{Meson decay.} 
    The production of dark photons in meson decays occurs mostly via transitions of the form
    \begin{align}
        X \to \gamma\Ap \to \gamma \phi\phi^{\dagger} \,,
        \label{eq:meson-decay}
    \end{align}
    where $X = \pi^0,\eta$. Reactions involving higher mass mesons are possible but usually subdominant, as shown for example in refs.~\cite{Berlin:2018pwi, SHiP:2020noy}. The $\rho$ and $\omega$ resonances, which lead to a sizeable enhancement of the production rate in a narrow mass region, are effectively taken into account within our formalism for proton bremsstrahlung, as detailed below. The transition in \cref{eq:meson-decay} can proceed either via an on-shell or an off-shell $\Ap$. We assume that the decay is dominated by the on-shell mode and make use of the formula~\cite{Gardner:2015wea, deNiverville:2011it}
    \begin{align}
        \frac{\text{BR}(X \to \gamma \Ap)}{\text{BR}(X \to \gamma \gamma)}
            \simeq 2 \, \epsilon^2 \, \bigg( 1 - \frac{\mAp^2}{m_{X}^2}\bigg)^3 \,,
            \qquad X = \pi^0, \eta
        \label{eq:decayrates}
    \end{align}
    for the corresponding branching ratio. This expression has been derived in the narrow width approximation (see ref.~\cite{Kahn:2014sra} for a discussion on the full treatment of off-shell and its implication on the sensitivity).
    
    As a first step towards computing the rate of DM--electron scattering events expected in the DUNE near detectors, we need to model the spectra of $\pi^0$s and $\eta$s produced in DUNE's primary target. This is crucial as any systematic bias in these spectra will propagate through the simulation chain and affect the yield of signal events. This is even more relevant if one considers, as done in this work, the possibility offered by the DUNE-PRISM concept to have a movable detector. We consider two samples of mesons:
    \begin{itemize}
        \item \emph{primary-only (SoftQCD)}, which includes only the mesons produced in the primary interaction of the proton impinging on the target, assuming that all protons eventually interact there. We employ {\sc Pythia}~v8.230~\cite{Sjostrand:2014zea} and adopt the flag {\tt SoftQCD} as the main mode for the simulation. For the simulation of relatively low-energy fixed target experiments with {\sc Pythia}, the flag {\tt SoftQCD:All} should be preferably used, as reported also in ref.~\cite{Berryman:2019dme}.
        
        \item \emph{beam-dump}, provided as ancillary material of ref.~\cite{Celentano:2020vtu}, which includes the interaction of secondary particles propagating in the DUNE target using {\sc Geant4}~\cite{Agostinelli:2002hh}.
    \end{itemize}
    The \emph{beam-dump} sample is more realistic and is taken as our default choice to study the DUNE sensitivity. We consider the other sample for sanity checks and for comparison with the result of ref.~\cite{DeRomeri:2019kic}, where {\sc Pythia} was employed for the simulation of mesons as well. We found discrepancies with the latter that we traced back to a different setup in the {\sc Pythia} generator.\footnote{We thank the authors of ref.~\cite{DeRomeri:2019kic} for helping us understand the {\sc Pythia} flags that were used in their simulation.} Further details and technical aspects on the comparison are reported in \cref{sec:DPcmp-deromeri}. 
    
    With the meson fluxes in hand, we use {\sc MadDump}~\cite{Buonocore:2018xjk} to generate samples of signal events of DM--electron scattering in the DUNE near detectors for all relevant parameter points. The program, once instrumented with a suitable UFO model file which encodes the coupling structure of the DM and the dark photon \cite{Degrande:2011ua}, takes care of all the steps of the simulation chain, from the decay of the parent mesons into DM particles to the detection process in the DUNE near detectors, including the computation of geometric acceptances and other effects due to the finite size of the detector.

    \begin{figure}
        \includegraphics[scale=0.9]{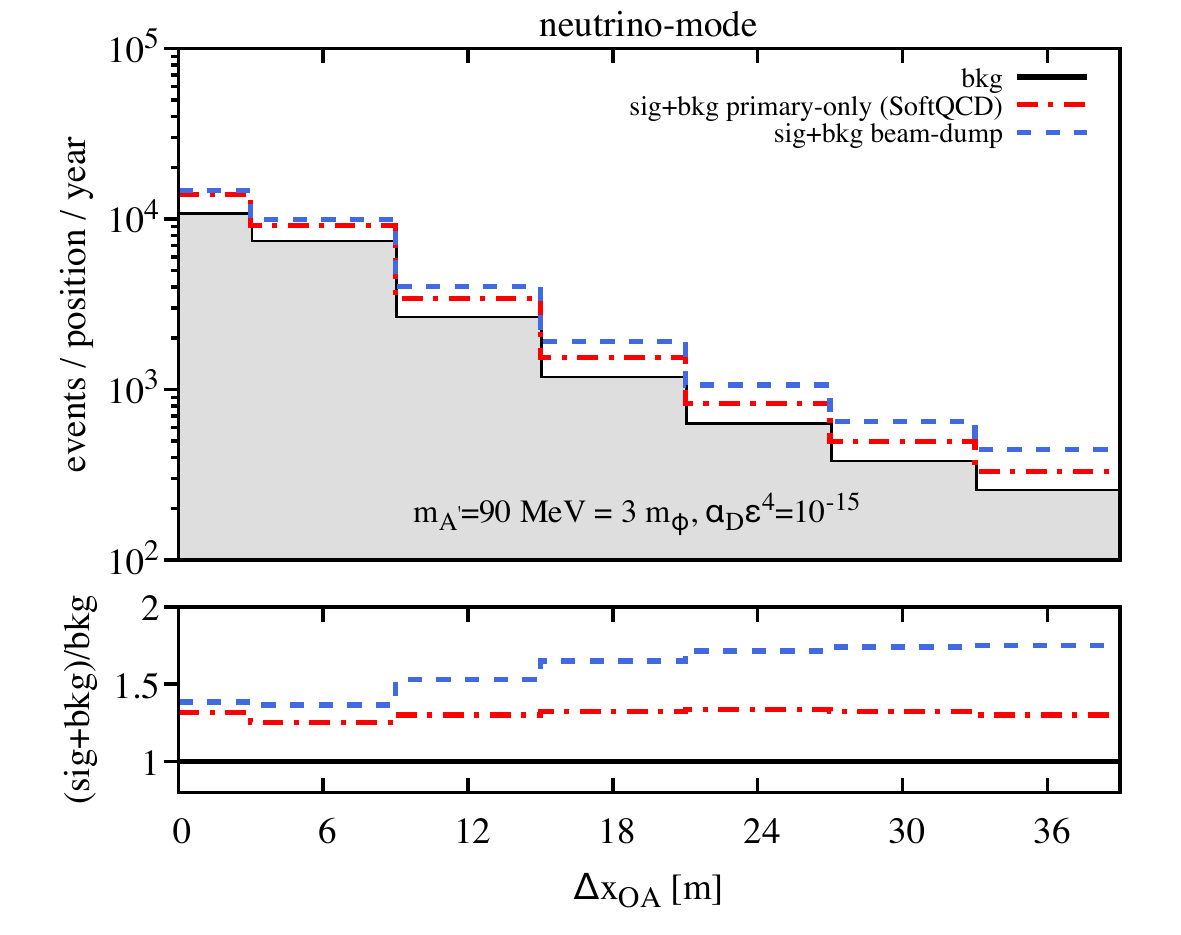}
        \caption{Expected number of DM--electron scattering events in the DUNE near detectors as a function of the detector position relative to the beam axis for the neutrino mode run. The background (solid gray histogram) corresponds to elastic neutrino and anti-neutrino scattering on electrons. Following ref.~\cite{DeRomeri:2019kic} we assume that the background from charged current quasi-elastic interactions can be made negligible by applying an energy-dependent cut on the lepton angle, which barely affects the signal and the elastic neutrino--electron scattering events. 
        For the signal\,+\,background histograms, we show separately the two signal samples introduced in the main text: the {\sc Pythia}-based sample of primary mesons, \emph{primary-only (SoftQCD)} (red dot-dashed), and the sample \emph{beam-dump} (blue dashed) based on the ancillary material of ref.~\cite{Celentano:2020vtu}.  See text for a discussion of the differences between these two samples.}
        \label{fig:mesons-cmp}
    \end{figure}

    In \cref{fig:mesons-cmp} we compare the results of the two different signal predictions by plotting the rate of signal\,+\,background events in each case as a function of the detector location $\Delta x_\text{OA}$ relative to the beam axis. The broad features visible in the figure are consistent with expectations: both the signal and background rates decrease as the detector is moved away from the focus direction of the beam.  The signal-to-background ratio, however, generally increases with $\Delta x_\text{OA}$ because the nearly massless neutrinos that are responsible for the background inherit more of the forward boost of their parent mesons than the much heavier $\Ap$ bosons from which the DM originates. Comparing our \emph{primary-only (SoftQCD)} (red dot-dashed) sample to the \emph{beam-dump} one (blue dashed), we observe that they are of similar size in the on-axis bins, but diverge with increasing off-axis angle: the event rate predicted by the \emph{beam-dump} sample drops less rapidly as the detector is moved away from the beam axis. This is indeed consistent with our expectation as there is a strong correlation between the location of the detector and the energy spectrum of the DM particles, which in turn originates from the spectra of the parent mesons. In particular, when the detector is on-axis, the flux it receives is dominated by the decay products of relatively energetic mesons, which are most likely produced in the primary proton--proton interaction. Therefore, this flux is reliably modeled by the \emph{primary-only (SoftQCD)} method.  Off-axis, however, the DM energy spectrum is dominated by much lower-energy particles, making it much more sensitive to the decays of soft mesons produced in secondary interactions in the target. The \emph{beam-dump} sample includes these mesons, explaining why the discrepancy between the blue and red curves in \cref{fig:mesons-cmp} increases with $\Delta x_\text{OA}$. We conclude that the modeling of secondary interactions is crucial for the off-axis strategy.

\paragraph*{Proton bremsstrahlung.} 
    In the dark photon mass range $\SI{500}{MeV} \lesssim \mAp \lesssim \SI{1}{GeV}$, i.e.\ above the $\eta$ threshold, proton bremsstrahlung dominates  dark photon production. Bremsstrahlung is preferentially emitted in the forward direction (collinear with the incoming proton) and in this limit can be well described by a generalization of the Fermi--Williams--Weizsäcker method~\cite{Fermi:1924tc, Williams:1934ad, vonWeizsacker:1934nji} (or ``equivalent photon method''). This method is based on the assumption that proton--nucleon scattering is dominated by exchange of vector bosons that are close to on-shell. Again we rely on {\sc MadDump} for our simulations, which implements bremsstrahlung following refs.~\cite{Blumlein:2013cua, PhysRevD.95.035006}. Let us parameterize the 4-momentum vector of the emitted $\Ap$ as $p_{\Ap} = (E_{\Ap}, p_\text{T} \cos(\phi), p_\text{T} \sin(\phi), zP)$,  with $E_{\Ap} \simeq z P + (p_\text{T}^2 + \mAp^2)/(2zP)$. Here, $P$ is the momentum of the incident proton, $z$ is the fraction of the proton momentum carried by the outgoing $\Ap$, $p_\text{T}$ is the momentum perpendicular to the beam momentum, and $\phi$ is the azimuthal angle. We generate unweighted $\Ap$ events according to the differential production rate
    \begin{align}
	    \frac{\mathrm d^2 N_{\Ap}}{\mathrm dz \,\mathrm dp^2_\text{T}} = \frac{\sigma_{pA}(s^\prime)}{\sigma_{pA}(s)}
	                                              F^2_{1,p}(\mAp^2)
	                                              w_{ba}(z,p^{2}_{\mathrm{T}}) \,,
        \label{eq:bremrate}
    \end{align}
    where $\sigma_{pA}(s)$ denotes the total interaction cross section of the incoming protons with a target nucleus of mass number $A$, $s=2m_p E_p$ is the square of the center-of-mass energy, and $s^\prime = 2 m_p (E_p - E_{\Ap})$. The ratio of cross sections $\sigma_{pA}(s^\prime) / \sigma_{pA}(s)$ compares the probability that the incoming proton interacts after having emitted a photon to its total interaction probability. For the proton form factor $F_{1,p}(\mAp^2)$, we use the parameterization from ref.~\cite{Faessler:2009tn} in the time-like region, so that off-shell mixing with vector mesons such as the $\rho$ and $\omega$ is effectively included in our calculation, leading to a resonance peak in the $\Ap$ production rate around $\mAp \simeq \SI{770}{MeV}$~\cite{Morrissey:2014yma}. Finally, the photon splitting function is
    \begin{multline}
	    w_{ba}(z,p^2_{\mathrm{T}}) =
	        \frac{\epsilon^2 \alpha}{2\pi H} \bigg[ \frac{1 + (1-z)^2}{z}
	                         - 2 z (1-z) \bigg( \frac{2 m_p^2 + \mAp^2}{H}
	                                          - z^2 \frac{2 m_p^4}{H^2} \bigg) \\
	                         + 2 z (1-z)[1 + (1-z)^2] \frac{m_p^2 \mAp^2}{H^2}
	                         + 2 z (1-z)^2 \frac{\mAp^4}{H^2} \bigg] \,,
	    \label{eq:wba}
    \end{multline}
    with $H = p^2_\text{T} + (1 - z) \mAp^2 + z^2 m_p^2$. The number of produced $\Ap$ events is estimated by integrating \cref{eq:bremrate} over a region well within the realm of the collinear approximation, i.e.\ a region where the kinematic conditions
    \begin{align}
        E_p, \, E_{\Ap}, \, E_p - E_{\Ap} \gg m_p, \mAp, p_{\mathrm{T}}
        \label{eq:bremcond}
    \end{align}
    hold.  In particular, following refs.~\cite{Blumlein:2013cua, PhysRevD.95.035006, Gorbunov:2014wqa}, we use the integration intervals $z \in [0.1,0.9]$ and $p_\text{T} < \SI{1}{GeV}$.

\subsection{Backgrounds}
\label{sec:dp-backgrounds}

As the experimental signature of $\phi$--$e^-$ scattering is a single, energetic recoil electron, the main backgrounds to this search in a neutrino beam experiment are due to neutrino--electron scattering and, if the final state hadronic system stays unidentified, charged current (CC) $\nu_e$--nucleon interactions. We estimate the backgrounds using GENIE~v3.00.06~\cite{Andreopoulos:2009rq}. In particular, we simulate the relevant processes -- $\nu$--$e^-$ scattering and CC $\nu_e$ interactions on nuclei -- in argon. The resulting event rates are then weighted bin-by-bin by the various on- and off-axis fluxes. The simulated DUNE neutrino fluxes have been taken from ref.~\cite{DUNEfluxes}, and extrapolated to higher energies, where Monte Carlo statistics in the simulations from ref.~\cite{DUNEfluxes} is too low. We perform this extrapolation by fitting the tail of the available fluxes linearly in log-space. The expected flux at very large neutrino energies, where the existing fluxes lack statistics, is then obtained by evaluating the fit function.

From the kinematics of elastic scattering, we know that both the signal process $\phi e^-\to\phi e^-$ and the background process $\nu e^-\to\nu e^-$ obey $E_e\theta_e^2<2m_e$, with $E_e$ ($\theta_e$) being the final state electron's total energy (scattering angle). Neutrino--nucleon interactions, on the other hand, lead to larger scattering angles. Since the expected angular resolution of DUNE-PRISM is sufficient~\cite{Acciarri:2015uup}, we impose the above condition as a kinematic cut, which leaves us with a comparatively small number of events involving neutrino--nucleon interactions. We therefore neglect the latter and only consider the elastic neutrino--electron scattering contribution in our analyses.

\begin{figure}
  \centering
  \includegraphics{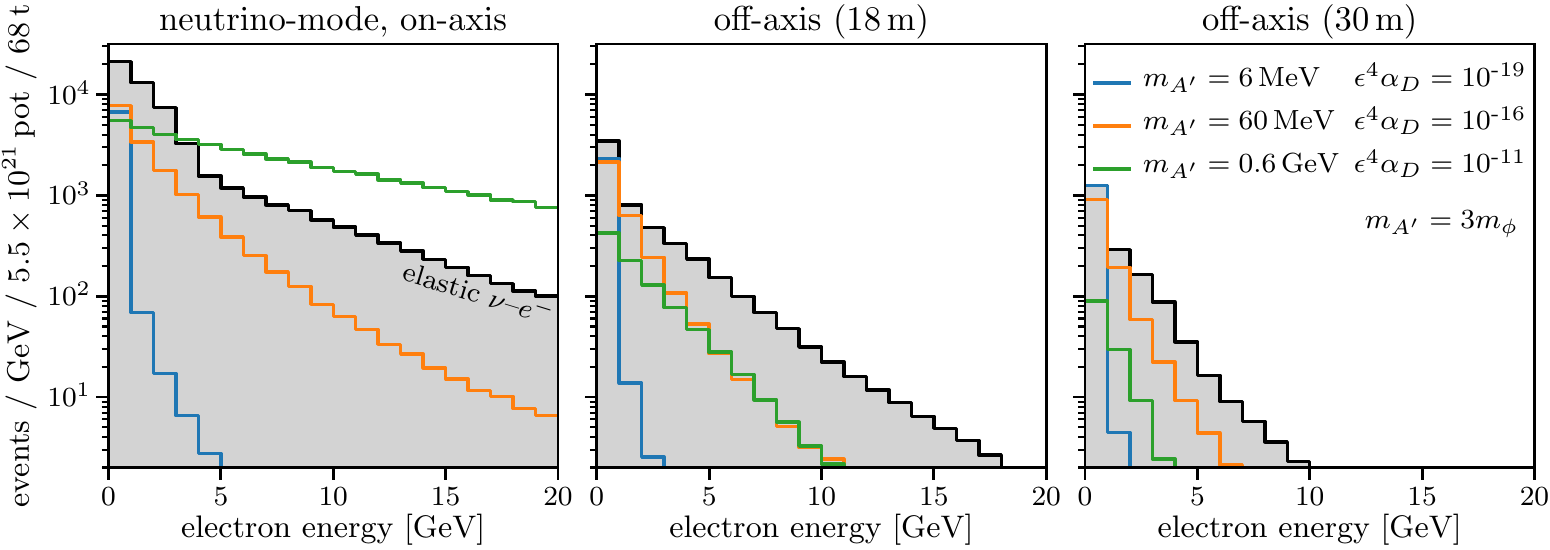}
  \caption{Signal and background spectra in DUNE-PRISM for the dark photon model after 5~years of data taking (\SI{5.5e21}{pot}) in neutrino mode.  Colored histograms show the DM--electron scattering signal for different exemplary sets of model parameter points, imposing $m_{A'} = 3 m_\phi$. The gray shaded background histogram is based on a simulation of elastic neutrino--electron scattering. Comparing data taken all on-axis (left panel) to data taken at \SI{18}{m} (\SI{31.36}{mrad}) off-axis (center panel) or \SI{30}{m} (\SI{52.26}{mrad}) off-axis (right panel), we see that, as expected, both the signal and background rates are significantly lower in the off-axis positions, especially at high energy.  At low energies, where most of the events are concentrated, the signal-to-background ratio becomes better when going off-axis.
  }
  \label{fig:darkphoton-rates}
\end{figure}

In \cref{fig:darkphoton-rates}, we plot the projected electron energy spectra for both the DM--electron scattering signal in the dark photon model and for the neutrino--electron scattering background.  We see that both the signal and the background peak at the lowest energies, but feature a long tail towards higher energies.  For the signal, the tail is most pronounced for heavier $A'$, corresponding to heavier DM particles, as we impose $m_{A'} = 3 m_\phi$ here. Heavy $A'$ are produced when heavy mesons decay in the LBNF target and are therefore only kinematically accessible in very hard proton--proton collisions.  The high-energy tails of both the signal and the background are strongly suppressed off-axis because the production of very energetic DM particles (for the signal) and neutrinos (for the background) occurs in production events that are strongly boosted in the forward direction.  Note that the signal for $m_{A'} = \SI{0.6}{GeV}$, $m_\phi = \SI{0.2}{GeV}$ (green histograms in \cref{fig:darkphoton-rates} drops of faster when going off-axis than the rates for lighter dark sectors.  This is because at $m_{A'} = \SI{0.6}{GeV}$ production is dominantly via bremsstrahlung, which is preferentially emitted in the forward direction.

\subsection{Statistical Analysis}
\label{sec:dp-statistics}

To derive sensitivity limits from our predicted signal and background rates, we use standard frequentist techniques. In particular, we test the signal\,+\,background hypothesis against simulated background-only data. To do so, we use a Poissonian log-likelihood function, $\log\mathcal{L}(\vec\Theta, \vec{X})$, defined in \cref{eq:likelihood}, which depends on the physical model parameters $\vec\Theta = (\epsilon^4\alpha_D, m_{A'}, m_\phi)$ and a set of nuisance parameters $\vec X$. The latter parameterize systematic normalization uncertainties and spectral ``tilts'' in both the signal and the background spectra. We consider systematic errors that are uncorrelated between different on-/off-axis positions (assuming 1\% relative error) in addition to errors which are correlated among all positions (assuming 10\% relative error). The sensitivity limits on $\epsilon^4\alpha_D$ for fixed values of $m_{A'}$ and $m_\phi$ are determined by comparing the log-likelihood ratio, defined in \cref{eq:Z-mu}, to the 90\% quantile of a $\chi^2$ distribution with one degree of freedom. See \cref{sec:statistics} for further details on our statistical procedure.

\subsection{Sensitivity to Light Dark Matter in the Dark Photon Model}
\label{sec:dp-sensitivity}

\begin{figure}
  \includegraphics{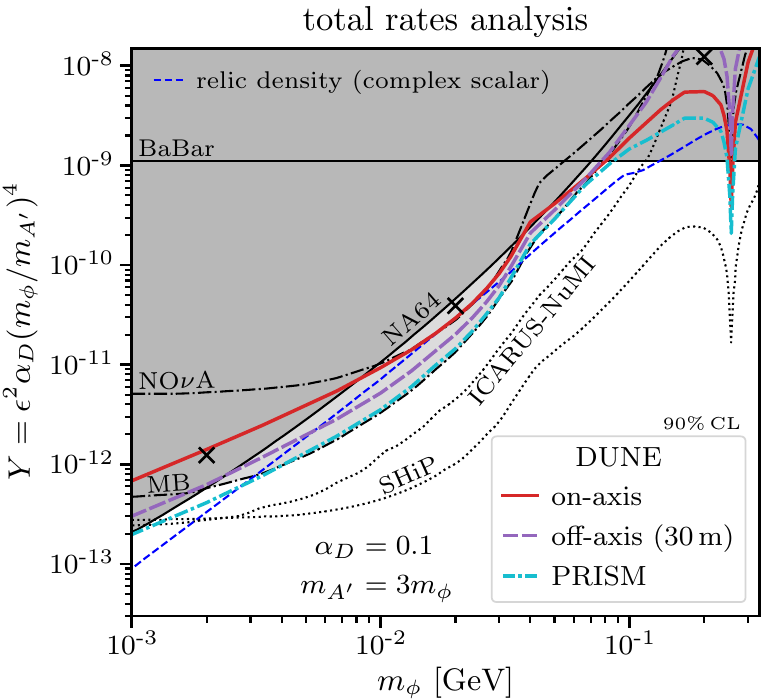}\hfill
  \includegraphics{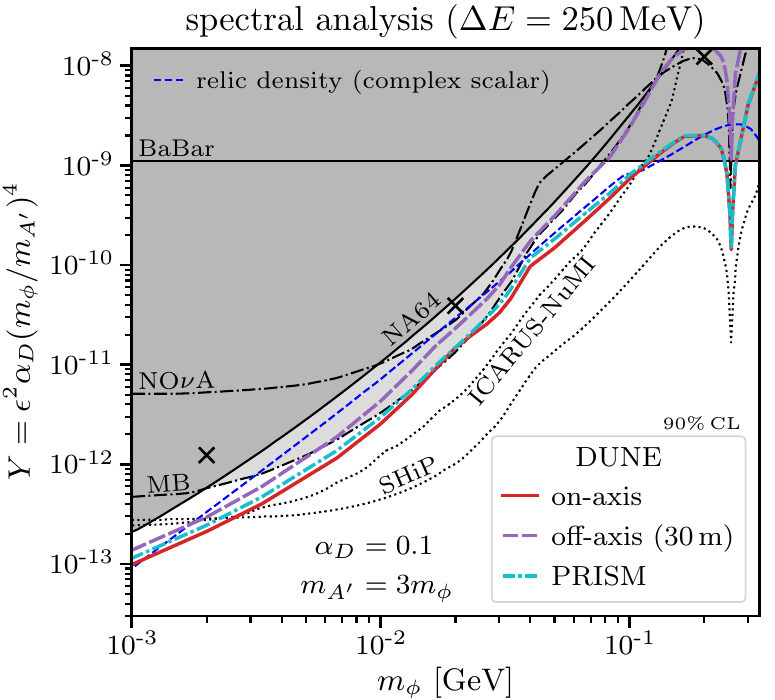}
  \caption{Expected upper exclusion limits for the dark photon  model, assuming a total DUNE-PRISM running time of 5~years ($\SI{5.5e21}{pot})$ in neutrino mode. We compare results for on-axis-only running (red solid), off-axis-only running (purple dashed), and a realistic DUNE-PRISM strategy with equal amounts of data taken at seven different locations (cyan dot-dashed). The left panel corresponds to a total rates (cut \& count) analysis as in ref.~\cite{DeRomeri:2019kic}, while the analysis in the right panel includes spectral information. Gray shaded regions indicate existing limits on the model, while dotted black lines show projections for other future experiments. Black crosses indicate the exemplary model parameter points presented in \cref{fig:darkphoton-rates}.
  }
  \label{fig:dp-limits}
\end{figure}

We present our main results for the dark photon model in \cref{fig:dp-limits}. In this figure, we compare on the one hand different running strategies: all data taken on-axis, all data taken off-axis, and combining data taken at different on-axis and off-axis locations as in DUNE-PRISM. On the other hand, we also compare two different analysis strategies, namely a total rates analysis ($n_\text{bins} = 1$ in \cref{eq:likelihood}) in the left panel and a spectral analysis ($n_\text{bins} = 80$ equal-width bins up to \SI{20}{GeV}) in the right panel.  The former type of analysis is similar to the one discussed in ref.~\cite{DeRomeri:2019kic}, and we confirm the main conclusion of these authors, namely that the DUNE-PRISM strategy of combining runs in seven different on-axis and off-axis locations (cyan dot-dashed) benefits the sensitivity to light DM in the dark photon model. It yields better results than both an on-axis-only run (red solid) and an off-axis-only run (purple dashed).  Interestingly, though, we reach a different conclusion when including the event spectrum: as shown in the right panel of \cref{fig:dp-limits}, the sensitivity in this case is about the same for on-axis-only running and for the PRISM strategy.  This can be understood by going back to \cref{fig:darkphoton-rates}, where we see that the signal-to-background ratio at energies $\gtrsim \SI{2}{GeV}$ is significantly better on-axis than it is off-axis. For a total rates analysis, these high energy events do not contribute because of the steep drop of the event rate compared to the lowest energy bins.  A spectral analysis, however, is able to harness also the statistical power of high-energy events and therefore suffers more from the poorer signal-to-background ratio in the off-axis position.

The overall shape of the exclusion curves in \cref{fig:dp-limits} can be understood mostly from the $\phi$ production rate, which drops at larger masses, where fewer production modes are available. The spectral feature at $m_\phi \sim \SI{230}{MeV}$, corresponding to $m_{A'} \sim \SI{700}{MeV}$ is related to the $\rho$ resonance: when $m_{A'} = m_\rho$, dark photons and $\rho$ vector mesons exhibit maximal mixing, leading to very efficient $A'$ production and thus strong limits.

We compare the DUNE sensitivity to existing limits from BaBar \cite{Lees:2017lec} and NA64 \cite{NA64:2019imj}, to a recast of NuMI off-axis data from NO$\nu$A \cite{deNiverville:2018dbu} and MiniBooNE (MB) \cite{Buonocore:2019esg}, and to the expected sensitivities of ICARUS-NuMI off-axis~\cite{Buonocore:2019esg} and SHiP \cite{SHiP:2020noy}. Note that we have rescaled the ICARUS-NuMI limit to an integrated luminosity of \num{2.5e21} protons on target (pot), corresponding to 5~years of NuMI running at the nominal beam power of \SI{700}{kW}.  We choose to present here some recasts that have not been officially approved by the respective experimental collaboration and for this reason have been omitted in some previous studies such as the ``Physics Beyond Colliders'' study at CERN \cite{Beacham:2019nyx}. We include such unofficial recasts only for neutrino experiments like NO$\nu$A \cite{deNiverville:2018dbu}, but not for other experiments like E137 \cite{Batell:2014mga,Celentano:2020vtu} or BEBC \cite{Buonocore:2019esg}. Also we do not present projections for future proposed experiments other than SHiP for the same reason, for a summary see ref.~\cite{Beacham:2019nyx}.  We see that DUNE-PRISM can probe important new regions of parameter space, with a sensitivity to $Y = \epsilon^2 \alpha_D(m_\phi/m_{A'})^4$ that is up to half an order of magnitude better than existing constraints for some $m_\phi$. Moreover, DUNE-PRISM can improve on the projected sensitivity for ICARUS-NuMI both in the small mass region ($m_{A'} \lesssim \SI{10}{MeV}$) and at large dark photon mass ($m_{A'} \gtrsim \SI{100}{MeV}$).

\section{Leptophobic Dark Matter}
\label{sec:leptophobic}

The second scenario we are going to consider in this paper is a leptophobic DM model, where the visible and the dark sector interact via a new gauge interaction under which the leptons are neutral. As a concrete example, we will consider the force associated with a gauged baryon number symmetry $U(1)_B$. We will call the leptophobic gauge field $Z'$ and the corresponding gauge coupling $g_Z$. The relevant terms in the Lagrangian of the leptophobic model are then
\begin{align}
  \mathcal{L}_\text{leptophobic}
    \supset 
      i g_Z z_{\phi}Z^{\prime \mu} J_{\mu}^{\phi}+ \partial_{\mu} \phi^\dagger \partial^{\mu} \phi - m_{\phi}^2 \phi^\dagger \phi    + g_Z z_q  \sum_q \bar{q} \gamma^\mu q \, Z'_\mu \,,
  \label{eq:L-leptophobic}
\end{align}
where $ J_{\mu}^{\phi} = \left[ (\partial_\mu \phi^\dagger) \phi - \phi^\dagger  (\partial_\mu \phi \right)]$. The sum in the last term runs over all quark flavors, and $z_\phi$, $z_q$ denote the $U(1)_B$ charges of DM and of SM quarks, respectively.  The leptophobic nature of the $Z'$ makes this particle rather elusive to experimental probes, as we will see in the following section. However, strong theoretical constraints arise from the fact that the $U(1)_B$ gauge symmetry is anomalous, so that extra fermions need to be added to the model to cancel anomalies~\cite{FileviezPerez:2011pt, Duerr:2013dza, anomaly1, FileviezPerez:2014lnj, anomaly2, Michaels:2020fzj}. Since these constraints are somewhat dependent on the ultraviolet completion of the model, we will not include them in our sensitivity plots.

\subsection{Dark Matter Production and Detection}
\label{sec:leptophobic-production}

We focus on $\Zp$ candidates with masses $\mZp \ge 2\GeV$. In this range, the dominant production mechanism is given by prompt production, which can be described by standard methods in perturbative QCD. 

The number of DM particles produced in the collision of primary protons impinging on the carbon target is given by the formula  
\begin{align}
    N_{\phi} = 2 N_\text{POT} \frac{\sigma_{pA\to\phi\phi^\dagger}}{\sigma_{pA,\text{tot}}}
             = A^{0.29} N_\text{POT}
               \frac{\sigma_{pN\to\phi\phi^\dagger}}{\sigma_{pN,\text{tot}}}
\end{align}
where $N_\text{POT}$ is the number of protons on target, $A = 12$ is the mass number of carbon (the target material used in the LBNF), and $\sigma_{pA}$ and $\sigma_{pN}$ stand for the proton--nucleus and proton--nucleon cross sections, respectively. In the above, we have assumed linear scaling with $A$ for the cross section of DM pair production, $\sigma_{pA \to \phi\phi^{\dagger}} = A \sigma_{pN\to\phi\phi^{\dagger}}$, and an effective total cross section per nucleon $\sigma_{pA,\text{tot}} = A^{0.71} \sigma_{pN,\text{tot}}$ for proton--carbon collisions~\cite{Abgrall:2011ae}. We adopt the approximation $\sigma_{pN,\text{tot}} = \SI{40}{mb}$~\cite{pdg2018} and compute $\sigma_{pA \to \phi\phi^{\dagger}}$ numerically in {\sc MadDump}, using a UFO implementation of the leptophobic model as done in ref.~\cite{Buonocore:2018xjk}. The computation is carried out at tree-level in QCD according to the standard parton model formula
\begin{align}
    \sigma_{pA \to \phi\phi^{\dagger}} = \sum_{a,b} \int \! \mathrm dx_1 \, \mathrm dx_2 \,
        f_{a/h_1}(x_1) \, f_{b/h_2}(x_2) \, \hat{\sigma}_{pA\to\phi\phi^{\dagger}}^{(a,b)} \,,
\end{align}
where $\hat{\sigma}_{pA\to\phi\phi^\dagger}^{(a,b)}$ is the partonic cross section for DM production in the scattering of two partons $a$ and $b$, and $f_{a/h_1}$ ($f_{b/h_2}$) are the corresponding parton distribution functions of parton $a$ ($b$) within hadron $h_1$ ($h_2$). In our simulation, we employ the leading order PDF set \texttt{NNPDF2.3LO}~\cite{Ball:2012cx,Ball:2013hta}, and we fix the factorization scale to $\mu_{F} = \mZp$. In the following, we will consider as a benchmark a scalar DM candidate $\phi$ of mass $m_\phi = \SI{750}{MeV}$ and $U(1)_B$ charge $z_\phi = 3$, while we vary the mass of the force carrier in the range $\mZp \in [2,7]\,\si{GeV}$. In this case, the branching fraction $\BR(\Zp \to \phi\phi^{\dagger})$ is of $\mathcal{O}(1)$~\cite{beam1} and, throughout this work, we assume it to be equal to 1.

Detection of leptophobic DM occurs via scattering of DM particles with the nuclei in the detector. Our search strategy focuses on the deep-inelastic scattering (DIS) signature, which is the most relevant one at multi-GeV energies. As for the dark photon model discussed in \cref{sec:dark-photon}, we expect better signal/background discrimination power when going off-axis~\cite{beam1}. The number of signal events is computed according to the formula
\begin{align}
    N_\text{sig} = \int\!\mathrm dz \, \rho(z) \int_{S}\!\mathrm dS \int\!\mathrm dE \,
                   \frac{\mathrm d^2N_{\phi}(E,S)}{\mathrm dE\,\mathrm dS}
                   \sigma_{\phi(\phi^{\dagger}) N\to\phi(\phi^{\dagger}) N} \,,
\end{align}
where $z$ denotes the coordinate along the beam axis, $\rho(z)$ is the number density of nucleons in the detector, $S$ is its surface orthogonal to the $z$-direction, and $\sigma_{\phi (\phi^{\dagger}) N \to \phi (\phi^{\dagger}) N}$ is the deep-inelastic DM--nucleon scattering cross section. The doubly differential flux ${\mathrm d^2N_{\phi}(E,S)}/(\mathrm dE\,\mathrm dS)$ of DM particles per unit area $\mathrm dS$ and per unit energy $\mathrm dE$ is computed on-the-fly by {\sc MadDump}.

\begin{figure}
    \includegraphics[scale=1.1]{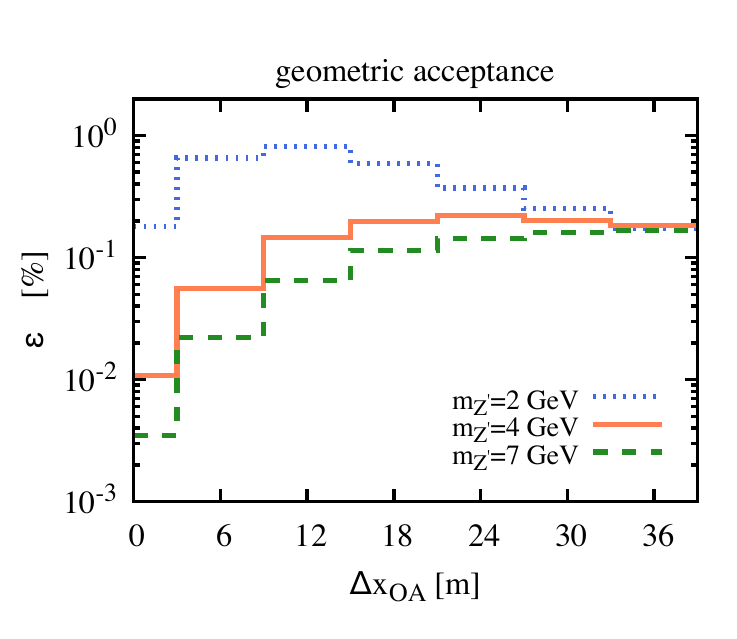} \hfill
    \caption{Geometric acceptance for leptophobic scalar DM particles as a function of the detector location relative to the beam axis. The different colored lines correspond to different masses of the $\Zp$ mediator. The drop of the acceptance when going towards the on-axis location is a consequence of the scalar nature of the DM candidate.}
    \label{fig:acceptance-lphobic_plus_bkg}
\end{figure}

In \cref{fig:acceptance-lphobic_plus_bkg} we show the geometric acceptance, which indicates what fraction of DM particles crosses the DUNE-PRISM detectors, as function of the off-axis location of the detectors for different masses $\mZp$ of the force carrier. We observe a distinctive pattern for higher masses: the acceptance steeply increases as the detectors are moved off-axis and then flattens at the largest off-axis angles, $\theta_\text{OA}$, attainable at DUNE-PRISM. This result is consistent with the ones shown in fig.~4 of ref.~\cite{beam1} and is a consequence of our DM candidate being a scalar. Indeed, this behavior can be understood by considering the differential production cross section for scalar DM particles, which reads ${\mathrm d\sigma}/{\mathrm d\theta_\text{OA}} \propto (1 - \cos^2\theta_\text{OA}) \sin\theta_\text{OA}$ in the center-of-mass frame. The suppression in the forward direction ($\theta_\text{OA} = 0$) is a consequence of angular momentum conservation.

Because of the increasing acceptance at large off-axis angles, combined with the expectation of lower neutrino-induced backgrounds far away from the beam axis, we consider not only the off-axis angles attainable in DUNE-PRISM, but also a hypothetical detector location at $\Delta x_\text{OA} = \SI{60}{m}$ ($\theta_\text{OA} = \SI{104.15}{mrad}$). This distance is similar to the one of the ICARUS-NuMI detector relative to the NuMI beam. According to the investigations carried out in ref.~\cite{beam2}, we expect that going that far off-axis leads to close-to-optimal sensitivity as it reduces the background to merely a few thousand events for an exposure of about $2.5\times 10^{21}$ pot. While exploiting off-axis distances as large as \SI{60}{m} in DUNE would require significant civil construction, doing so might be interesting if hints for leptophobic DM interactions should be found in DUNE-PRISM. We use {\sc MadDump} to compute the simulated deep-inelastic neutrino--nucleon scattering events within the detector. For reference, in \cref{tab:lphopbic_additional_setups} we collect the main parameters of different DUNE-PRISM configurations as well as ICARUS-NuMI. In simulating ICARUS-NuMI, we have used the NuMI flux from \cite{Adamson:2008qj}.

\begin{table}
    \begin{tabular}{lC{1.3cm}C{1.1cm}cC{2.2cm}C{1.75cm}C{1.8cm}C{1.8cm}}
        \toprule
        Experiment & Baseline & Mass & Off-axis & Exposure & $n_\text{bkg}$
                   & $\varepsilon_{\mZp=2\GeV}$ & $\varepsilon_{\mZp=7\GeV}$  \\
                   &   [m]    &  [t] &  [mrad]  &   [pot]  & & & \\\midrule
        DUNE on-axis        & 574 & \phantom{4}68 & \phantom{104.1}0 & \num{5.5e21} & \num{2.6e7} & \num{1.8e-3} & \num{3.5e-5} \\
        DUNE\,@\,\SI{30}{m} & 574 & \phantom{4}68 & \phantom{1}52.26 & \num{5.5e21} & \num{2.9e5} & \num{2.5e-3} & \num{1.6e-3}\\
        DUNE\,@\,\SI{60}{m} & 574 & \phantom{4}68 & 104.15           & \num{5.5e21} & 3950        & \num{6.3e-4} & \num{1.0e-3}\\
        ICARUS-NuMI         & 789 & 480           & \,$\sim 100$     & \num{2.5e21} & 1600        & \num{1.5e-4} & \num{2.7e-4}\\
        \bottomrule
    \end{tabular}
    \caption{Summary of experimental setups considered for leptophobic DM. We have used a primary proton energy of \SI{120}{GeV} and the expected exposure after 5~years for both DUNE and ICARUS-NuMI. A cut on the visible energy ($E_\text{vis} > \SI{3}{GeV}$) is applied to all signal and background events, with $n_\text{bkg}$ being the number of background events that pass the cut. The two rightmost columns indicate the geometric acceptance at two different $\Zp$ masses.}
    \label{tab:lphopbic_additional_setups}
\end{table}

\subsection{Backgrounds}
\label{sec:leptophobic-backgrounds}

As DM in the leptophobic model scatters only on hadrons, the main irreducible background channel is neutral current (NC) neutrino scattering. Charged current (CC) interactions might also lead to signal-like signatures when the final state charged lepton is misidentified as a charged pion. We expect the latter contribution to be smaller, as the NC and CC cross sections are of similar size, and hence we neglect it in the following.  

The expected background distributions are obtained as described in \cref{sec:dp-backgrounds}, i.e.\ by using GENIE to simulate events and then weighting the events with the appropriate on- and off-axis neutrino fluxes.
For studies at $\Delta x_\text{OA} = \SI{60}{m}$, we compute the neutrino flux based on DUNE's flux Ntuples published in ref.~\cite{DUNEfluxes}, while for smaller off-axis distances we use directly the flux histograms from the same reference.
Considering only data with $E_\text{vis} > \SI{3}{GeV}$ allows us to neglect any contributions from resonant scattering, quasi-elastic scattering, and coherent scattering, leaving us with NC deep-inelastic neutrino scattering as the only relevant background process.

\begin{figure}
    \includegraphics{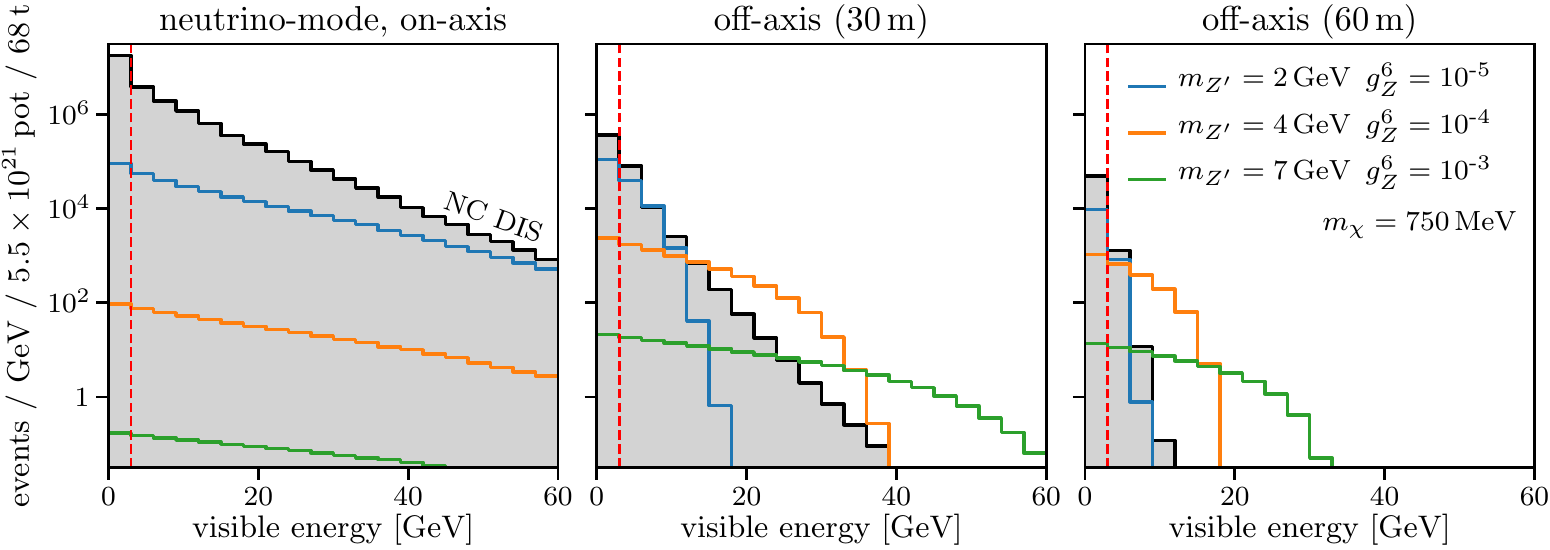}
    \caption{Predicted DM scattering signal in the leptophobic model and corresponding backgrounds for DUNE-PRISM with \SI{5.5e21}{pot} of neutrino-mode data, corresponding to 5 years of running.  Colored histograms show the signal rate for different exemplary model parameter points. The gray shaded background histogram is based on a simulation of NC deep-inelastic neutrino--nucleus scattering. The left panel is for the on-axis location, while the center and right panels correspond to data taken at \SI{30}{m} (\SI{52.26}{mrad}) off-axis and at a hypothetical detector location \SI{60}{m} (\SI{104.15}{mrad}) off-axis, respectively. The red dashed line indicates the energy threshold $E_\text{vis}>\SI{3}{GeV}$ imposed in our analysis. We observe that backgrounds are significantly suppressed off-axis, especially at high energies. The signal, on the other hand, is even enhanced there for large $Z'$ masses, as explained in \cref{sec:leptophobic-production}.}
    \label{fig:leptophobic-rates}
\end{figure}

The predicted signal and background spectra for the leptophobic model are shown in \cref{fig:leptophobic-rates}. As for the dark photon model (cf.\ \cref{fig:darkphoton-rates}), we find that backgrounds are dramatically suppressed when going off-axis. NC interactions with a given visible energy $E_\text{vis}$ are typically caused by neutrinos with an energy $E_\nu \gg E_\text{vis}$, implying that they are very sensitive to the high-energy tail of the neutrino spectrum. As the latter is strongly suppressed off-axis, so is the rate of NC background events to our DM search. Because of the lower backgrounds as well as the increased geometric acceptance off-axis (see \cref{fig:acceptance-lphobic_plus_bkg}), we observe a dramatic increase in the signal-to-background ratio, especially for $m_{Z'}$ in the multi-GeV range.

\subsection{Sensitivity to Leptophobic Dark Matter}
\label{sec:leptophobic-sensitivity}

\begin{figure}
    \includegraphics{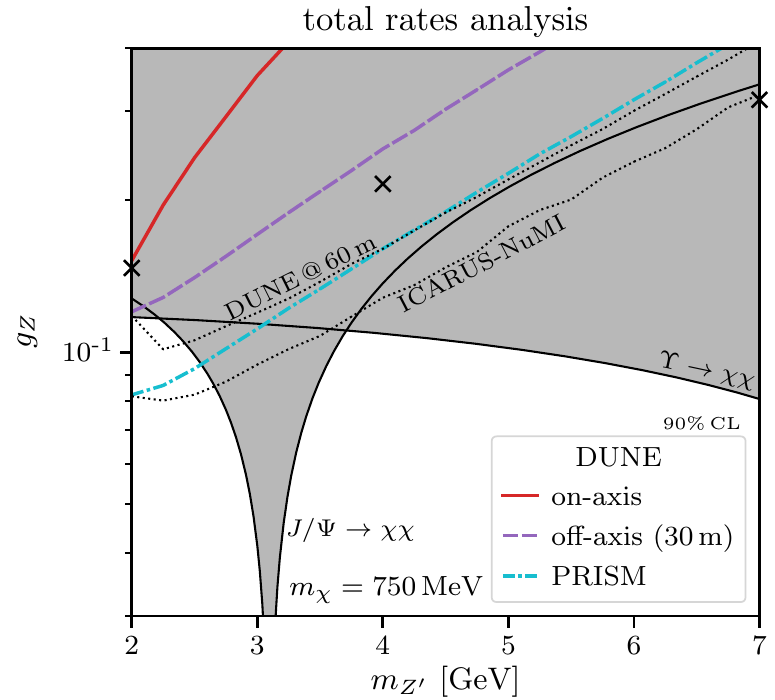}\hfill
    \includegraphics{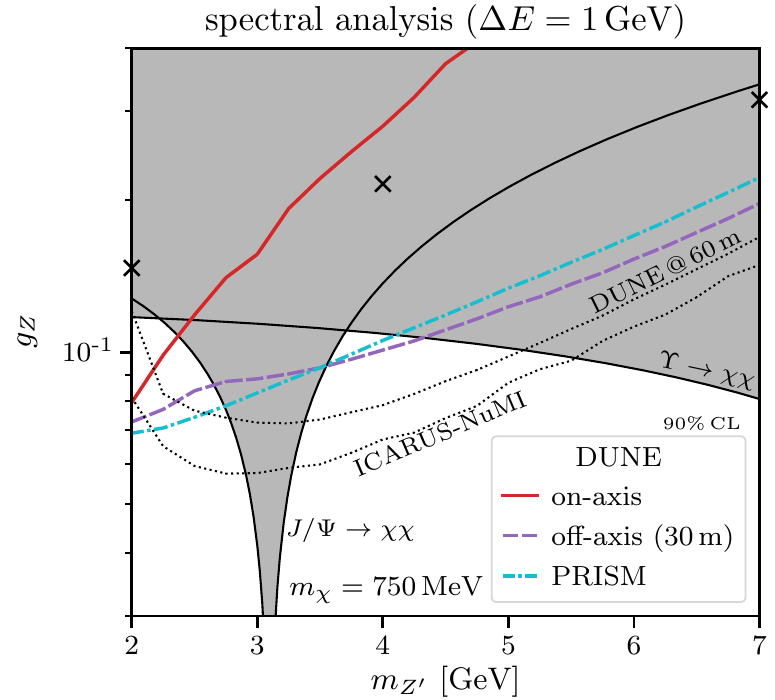}
    \caption{
    Projected upper exclusion limits for the leptophobic DM model, assuming the DM particle $\phi$ carries a $U(1)_B$ charge of $z_{\phi} = 3$. Colored exclusion curves correspond to different DUNE-PRISM running strategies namely on-axis-only running (red solid), off-axis-only running (purple dashed), and equal running times at seven different locations (cyan dot-dashed). Black dotted exclusion curves show the constraints that could be obtained at ICARUS-NuMI using neutrinos from the NuMI beam and at a hypothetical detector placed \SI{60}{m} off-axis in the DUNE/LBNF beam.  Like for the dark photon model (cf.\ \cref{fig:dp-limits}), we compare a total rates analysis (left panel) to an analysis harnessing the full energy spectrum of events (right panel). We also compare to existing limits from invisible $J/\psi$ and $\Upsilon$ decays at BaBar~\cite{Graesser:2011vj, Aubert:2009ae} (gray shaded regions). The black crosses indicate the exemplary model parameter points presented in \cref{fig:leptophobic-rates}.}
    \label{fig:leptophobic-limits}
\end{figure}

To estimate the sensitivity of DUNE-PRISM to leptophobic DM, we follow the same statistical procedure as in \cref{sec:dp-sensitivity}, employing in particular the log-likelihood ratio from \cref{eq:Z-mu,eq:likelihood} which depends on the model parameters $\vec\Theta = (g_Z^6, m_{Z'}, m_\chi)$ in this case. Our results are shown in \cref{fig:leptophobic-limits}, comparing once again a total rates (cut \& count) analysis ($n_\text{bins} = 1$, left panel) to an analysis utilizing spectral information ($n_\text{bins} = 57$ equal-width bins between \SI{3}{GeV} and \SI{60}{GeV}, right panel). We find that the spectral analysis clearly outperforms the total rates fit, as can be understood from the event spectra in \cref{fig:leptophobic-rates}. As for leptophilic DM (see \cref{sec:dp-sensitivity}), the DUNE-PRISM approach plays out its strengths especially for the total rates analysis. Unlike for leptophilic DM, however, on-axis only running is never optimal in the leptophobic model, no matter which type of analysis is used. An off-axis-only run would, however, be competitive with the DUNE-PRISM strategy if a spectral analysis is performed. This behavior can be understood from the better signal acceptance and lower backgrounds that we have discussed above in the context of \cref{fig:acceptance-lphobic_plus_bkg,fig:leptophobic-rates}. For the same reason, a hypothetical detector at \SI{60}{m} off-axis (upper black dotted line in \cref{fig:leptophobic-limits}) would outperform DUNE-PRISM at any of the available on-axis and off-axis locations. Nevertheless, ICARUS-NuMI will do even better using off-axis neutrinos from the NuMI beam. The excellent performance of ICARUS-NuMI can be attributed to its large mass of \SI{480}{t}, compared to only \SI{68}{t} for the combination of the DUNE-PRISM ND-LAr and ND-GAr detector. In addition, the NuMI neutrino flux drops off somewhat faster and becomes softer than the DUNE/LBNF flux when going off-axis.
Comparing to existing limits from invisible $J/\psi$ and $\Upsilon$ decays at BaBar~\cite{Graesser:2011vj, Aubert:2009ae}, we find that DUNE-PRISM will compete with these existing constraints only in small regions around $m_{Z'} \sim \SI{2}{GeV}$ and $m_{Z'} \sim \SI{4}{GeV}$. 

\section{Heavy Neutral Leptons}
\label{sec:hnl}

Heavy neutral leptons (HNLs), often also called sterile neutrinos, are $SU(3)_c
\times SU(2)_L \times U(1)_Y$-singlet fermions whose only (non-gravitational)
coupling to the SM is via neutrino mixing. The corresponding operator is
\begin{align}
  \mathcal{L}_\text{HNL} \supset y \bar{L} \tilde{H} N \,,
  \label{eq:L-HNL}
\end{align}
where $N$ is the HNL field (a Weyl fermion), $L$ are the left-handed lepton
doublets, $\tilde{H} = i \sigma^2 H^*$ is the conjugate of the SM Higgs doublet
field, and $y$ is a dimensionless Yukawa coupling.  Once the Higgs field acquires
a vacuum expectation value, the operator in \cref{eq:L-HNL} leads to mass mixing
between the HNL and the active neutrinos. The HNL thus acquires the
same couplings as the active neutrino and can be produced in any process that
produces neutrinos in the SM, unless forbidden by kinematics. Consequently,
given a large enough coupling $y$ in \cref{eq:L-HNL}, meson decays in the
DUNE target can copiously produce HNLs.

Feynman diagrams involving the mixing operator from \cref{eq:L-HNL} also
admit HNL decays into various final states involving neutrinos, charged
leptons, and/or hadrons.  If the HNL is sufficiently long-lived (that is,
if $y$ is not too large), some of these decays will occur inside the
near detector, leaving unique signatures.

The sensitivity of the DUNE near detectors to HNLs has been studied before
in ref.~\cite{Berryman:2019dme,Ballett:2019bgd}, albeit for an on-axis configuration only.
Current global constraints on HNLs have been compiled for instance in
refs.~\cite{Atre:2009rg, Drewes:2015iva, deGouvea:2015euy, Bolton:2019pcu}.

\subsection{Production of Heavy Neutral Leptons}
\label{sec:hnl-production}

In our simulations of HNL production and decay, we closely follow
ref.~\cite{Ballett:2019bgd}, and we use a modified and expanded version of the
{\sc NuShock} code~\cite{NuShock} accompanying ref.~\cite{Ballett:2019bgd} \footnote{ As it has been pointed out in \cite{Coloma:2020lgy} there are significant discrepancies in the computation of the HNL branching ratios.However, these discrepancies do not affect our conclusions. }. More
precisely, we predict the HNL flux in DUNE by starting from the simulated
neutrino production events released
by the DUNE collaboration~\cite{DUNEfluxes}, assuming a total luminosity
of $\SI{5.5e21}{pot}$, all taken in neutrino
mode. Each event in these files, which
are based on DUNE's full Monte Carlo simulation of the target station
and decay volume, corresponds to the production of a single SM neutrino.
To obtain HNL production events instead, we extract the kinematics of
the neutrinos' parent mesons from the Monte Carlo event files and then
use {\sc NuShock} to re-generate their decays, enforcing the production of an HNL
instead of an SM neutrino in the final state. {\sc NuShock} accounts for the
reduced branching ratio to the HNL final state by including an appropriate
weight factor for each event.

Unfortunately, the DUNE/LBNF simulation includes only neutrino production in
pion and kaon decays, but not in charm decays. This is understandable, given
that neutrinos from charm decay are completely irrelevant in the SM.  For HNL
searches, however, charm decays are very important because at HNL masses larger
than the kaon mass, they are the only kinematically allowed HNL production
channel.  We therefore estimate the flux of charm mesons following once again
refs.~\cite{Ballett:2019bgd,NuShock}. We estimate the charm production rate
based on the cross section $\sigma_{c\bar{c}} = \SI{2}{\micro b}$ for
$c\bar{c}$ (open charm) production in proton--proton scattering
at the center-of-mass energy $\sqrt{s} =
\sqrt{2 (\SI{1}{GeV}) \cdot (\SI{120}{GeV})} \simeq \SI{15}{GeV}$, see fig.~16 in ref.~\cite{Lourenco:2006vw}.  Taking the ratio of
$12 \sigma_{c\bar{c}}$ to the total proton--\iso{C}{12} cross section of
$\sigma_{p C} = \SI{331.4}{mb}$ \cite{RamanaMurthy:1975vfu} gives the rate of
charm production per pot. Multiplying further by the fragmentation fraction
into $D_s$, $f_{D_s} = 0.077$ \cite{Abramowicz:2013eja}, we obtain the rate of
$D_s$ production.  To generate the $D_s$ kinematics in the center-of-mass
frame, we approximate the momentum dependence of the differential $D_s$ production
cross section in the center-of-mass frame as \cite{Aoki:2017spj}
\begin{align}
  \frac{\mathrm d^2\sigma}{\mathrm dx\,\mathrm dp_{T,D_s}} \propto (1 - |x|)^n \exp(-b\,p_{T,D_s}^2) \,,
\end{align}
where $p_{T,D_s}$ is the $D_s$ transverse momentum and $x \equiv 2 p_{z,D_s}
/ \sqrt{s}$, with $p_{x,D_s}$ the $D_s$ momentum along the beam axis.
For the numerical coefficients, we use the values $n = 6.1$ and
$b = 1.08$~\cite{Aoki:2017spj}, which have been measured in the Fermilab E769 experiment
using a \SI{250}{GeV} proton beam impinging on a fixed target.

\begin{figure}
  \hspace*{-1.5cm}\includegraphics[width=1.2\textwidth]{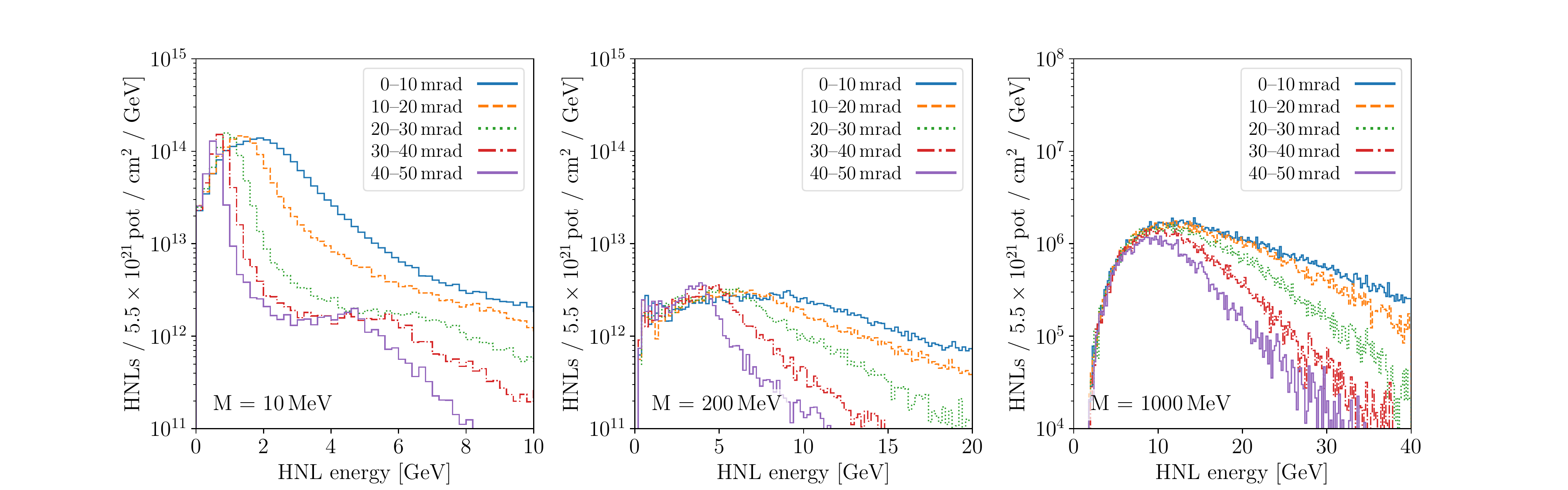}
  \caption{
    The differential flux of HNLs at the DUNE near detector
    site (\SI{574}{m} baseline) for different HNL masses (left, middle, and right
    panels) and for different off-axis angles (different colors and line
    styles). For illustrative purposes, we have made the unphysical assumption
    $|U_{e4}|^2 = |U_{\mu 4}|^2 = |U_{\tau 4}|^2 = 1$. In other words, we have
    assumed that all neutrino production processes in the DUNE/LBNF beamline
    are replaced by the corresponding HNL production processes.
    The fluxes shown here can therefore be linearly rescaled to smaller and
    more realistic values of $|U_{\alpha 4}|^4$.  Note the different vertical
    and horizontal axis scales in the three panels.}
  \label{fig:hnl-flux}
\end{figure}

Our predicted HNL flux as a function of HNL energy is shown in
\cref{fig:hnl-flux} for different HNL masses and different off-axis angles.
The left panel ($M = \SI{10}{MeV}$) corresponds to a practically massless HNL
that is dominantly produced in pion decays; the HNLs in the middle panel ($M =
\SI{200}{MeV}$) are too heavy to benefit from this production mode, and can
only be produced in kaon decays, which explains the much lower flux; at $M =
\SI{1}{GeV}$ (right panel), only $D$ meson decays contribute to HNL production.
In \cref{fig:hnl-flux}, we have assumed $|U_{e4}|^2 = |U_{\mu 4}|^2 = |U_{\tau 4}|^2 = 1$,
implying that all meson decays involving neutrinos in the SM are
replaced by the corresponding decays into HNLs.

Not surprisingly, the HNL spectrum becomes softer at larger off-axis angles
because higher energy parent mesons are more strongly boosted in the forward
direction.  This softening of the HNL spectrum when going off-axis is fully
analogous to the off-axis softening of the SM neutrino spectrum.  Both on-axis
and off-axis, the spectra of heavier HNLs (from kaon and charm decays) are
significantly harder than the spectra of light HNLs because the smaller boost
factors of heavy mesons render their decays more isotropic.

\subsection{Heavy Neutral Lepton Decay}
\label{sec:hnl-decay}

To determine the rate of HNL decays inside the DUNE near detectors, we follow
the trajectories of all simulated HNLs to determine which of them cross the detectors.
We consider both the segmented liquid argon time projection chamber
referred to as ArgonCube or ND-LAr by the DUNE collaboration and
the pressurized gaseous argon TPC (``multi-purpose detector'' or ND-GAr).
The ND-LAr detector is box-shaped, with a width (perpendicular to the beam axis)
of \SI{7}{m}, a height of \SI{3}{m}, and a depth (along the beam axis) of
\SI{5}{m}. The ND-GAr detector is cylindrical, with the
cylinder axis oriented horizontally and perpendicular to the beam axis.
The detector's width is \SI{5}{m}, and the cylinder radius is \SI{2.6}{m}
\cite{DUNE:2021tad}. We compute the 3D coordinates at which
each trajectory enters and exits the detectors by using some elementary
geometry, courtesy of the {\sc trimesh} Python package~\cite{trimesh}.
We then weight each HNL with the probability for decaying inside either
of the two detectors.

HNL lifetimes and decay branching ratios into different final states as
functions of mass are once again computed using {\sc NuShock}.  In setting limits,
we will consider in particular the following final states:
\begin{enumerate}
  \item $\nu e^+ e^-$ because it has the largest branching ratio among the
    visible decay channels for HNL masses below the muon threshold. The channel
    contributes appreciably also at higher masses, and, moreover,
    backgrounds are small in this channel.
  \item $\nu e^\pm \mu^\mp$ and $\nu \mu^+ \mu^-$ due to the large branching
    ratios for heavy HNLs as well as low background levels
  \item $\nu \pi^0$, $e^\pm \pi^\mp$ and $\mu^\pm \pi^\mp$, which dominate at
    intermediate HNL masses, but are also important at large masses. These
    channels suffer from large backgrounds that arise from NC or CC neutrino
    interactions with emission of an extra pion.
\end{enumerate}
For each decay channel, we have used {\sc NuShock} to simulate a large sample of HNL
decays at rest.  For each HNL in the DUNE/LBNF beam that decays inside the detector,
we randomly pick one of the decay events and boost it from the HNL rest frame
to the laboratory frame to obtain the 4-momenta of the observable decay products.

\begin{figure}
  \centering
  \includegraphics[width=0.8\textwidth]{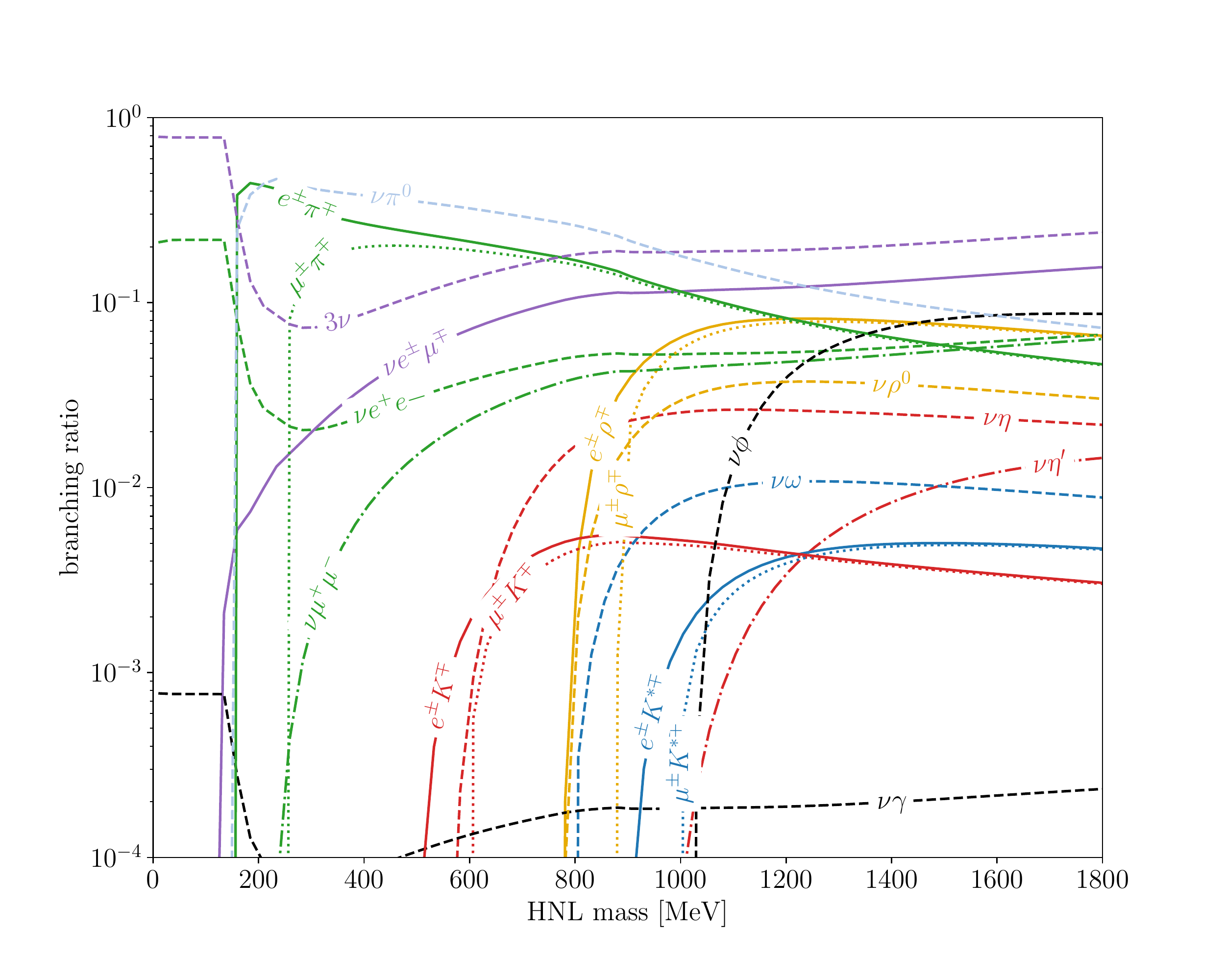}
  \caption{Branching ratios of HNL decay modes as a function
    of mass. We have assumed flavor-universal mixing, i.e.\ $|U_{e4}| = |U_{\mu
  4}| = |U_{\tau 4}|$.}
  \label{fig:hnl-br}
\end{figure}

The HNL decay  branching ratios are plotted in \cref{fig:hnl-br} as a function
of HNL mass (see also ref.~\cite{Ballett:2019bgd}).  We see that, at low HNL
mass $\lesssim m_\pi$, the dominant decay mode is invisible, $N \to 3\nu$,
followed by $N \to \nu e^+ e^-$.  Above the pion threshold, two-body final
states involving charged or neutral pions begin to dominate. Nevertheless,
fully leptonic three-body final states ($\nu e^+ e^-$, $\nu \mu^+ \mu^-$, $\nu
e^\pm \mu^\mp$) remain relevant, and $N \to \nu e^\pm \mu^\mp$ even becomes the
strongest visible decay mode at $m_N \gtrsim \SI{1.2}{GeV}$.  Decay modes
involving kaons, $\rho$ mesons, and other hadrons become important as the
corresponding kinematic thresholds are crossed.

\subsection{Backgrounds and Analysis Cuts}
\label{sec:hnl-backgrounds}

To estimate the backgrounds to an HNL search in the DUNE near detectors, we
have used GENIE~v3.00.06~\cite{Andreopoulos:2009rq} to simulate CC and NC
neutrino interactions on argon. We have simulated $\num{3e6}$ events per
flavor, one million each within each of the neutrino energy intervals $[0, 20]$\,GeV,
$[20, 40]$\,GeV, and $[40, 60]$\,GeV.  Within each interval, neutrino energies
are distributed according to the on-axis DUNE flux.  To
obtain off-axis event samples, events are appropriately re-weighted with the
ratio of the off-axis and on-axis fluxes. The rationale for dividing the simulation
into three different energy ranges is to obtain more Monte Carlo statistics at
high energies, where much of the sensitivity to HNLs is coming from.

We process the simulated events through the background simulation implemented
in {\sc NuShock} \cite{NuShock,Ballett:2019bgd}. This involves a very simple
detector simulation that applies Gaussian smearing to the momenta of the final state
particles and uses a kinetic energy threshold to determine which
of them are reconstructed. It rejects any events containing hadrons other than
pions or atomic nuclei, thus exploiting the fact that the HNL decay modes that we
consider do not contain heavy hadrons, while many potential background processes are from
deep-inelastic neutrino scattering events that typically involve a lot of
hadronic activity. Note that $\pi^0$s, which are by default not decayed in GENIE,
are instead decayed in {\sc NuShock} and the analysis is carried out on the two final
state photons. Charged pions, on the other hand, are retained undecayed due to
their much longer lifetime.

The next step consists of simple particle misidentification
rules. In particular:
\begin{itemize}
    \item Pions are misreconstructed as muons if their randomly chosen track length
    is sufficiently long ($> \SI{2}{m}$).
    \item $e^+ e^-$ pairs are misreconstructed as (converted) photons if their angular
    separation is below a threshold ($3^\circ$).
    \item Photons that convert after less than \SI{2}{cm} are reconstructed as
    electrons or positrons.
\end{itemize}
The list of final state particles after threshold cuts and misidentification
is then compared to the list of final state particles expected for the signal
channel under consideration. Only if the number and type of each particle
match between the signal and background, the event is retained (exclusive
analysis). We do, however, conservatively assume no charge identification
capabilities even though the ND-GAr detector will have a magnetic field and
should therefore be able to efficiently distinguish positively and negatively
charged final state particles.

To further suppress backgrounds, we exploit the fact that HNL decay products
are typically strongly boosted in the forward direction, while background
events have a more isotropic topology.
We implement a cut on the angle $\theta$ between the mean direction of the two visible would-be HNL decay products and the beam axis. For given HNL mass $M$, we set the threshold at $\theta < M / (E_1 + E_2)$, where $E_1$ and $E_2$ are the energies of the two would-be HNL decay products. This threshold value is an estimate for the forward boost of the decay products in a real HNL decay.

We find the dominant background contributions for the different HNL decay channels
to be as follows:
\begin{enumerate}
    \item For $N \to \nu e^+ e^-$, by far the most important background is due to
      misidentified photons.
    \item $N \to \nu \mu^+ \mu^-$ is most easily mimicked by CC $\nu_\mu$ interactions
      with one real muon and one misidentified charged pion.
    \item Similarly, $N \to \nu e^\pm \mu^\mp$ suffers from a background due to
      CC $\nu_\mu$ interactions with a real muon and a photon that is misidentified
      as an electron.
    \item Backgrounds to $N \to e^\pm \pi^\mp$ arise mostly from CC $\nu_e$
      interactions with pion production. A smaller contribution comes from NC neutrino
      interactions with a real pion and a photon misidentified as an electron.
    \item Similarly, $N \to \mu^\pm \pi^\mp$ is affected by a large background
      due to CC $\nu_\mu$ interactions with pion production.
    \item Finally, in $N \to \nu \pi^0$, the source of the large background is NC
      neutrino interactions with pion production.
\end{enumerate}
We have also considered neutrino trident production (which is not simulated by GENIE) as a possible source of background. However, based on the calculation in ref.~\cite{Ballett:2018uuc}, we conclude that this background will be negligible.

\begin{figure}
  \centering
  \includegraphics[width=\textwidth]{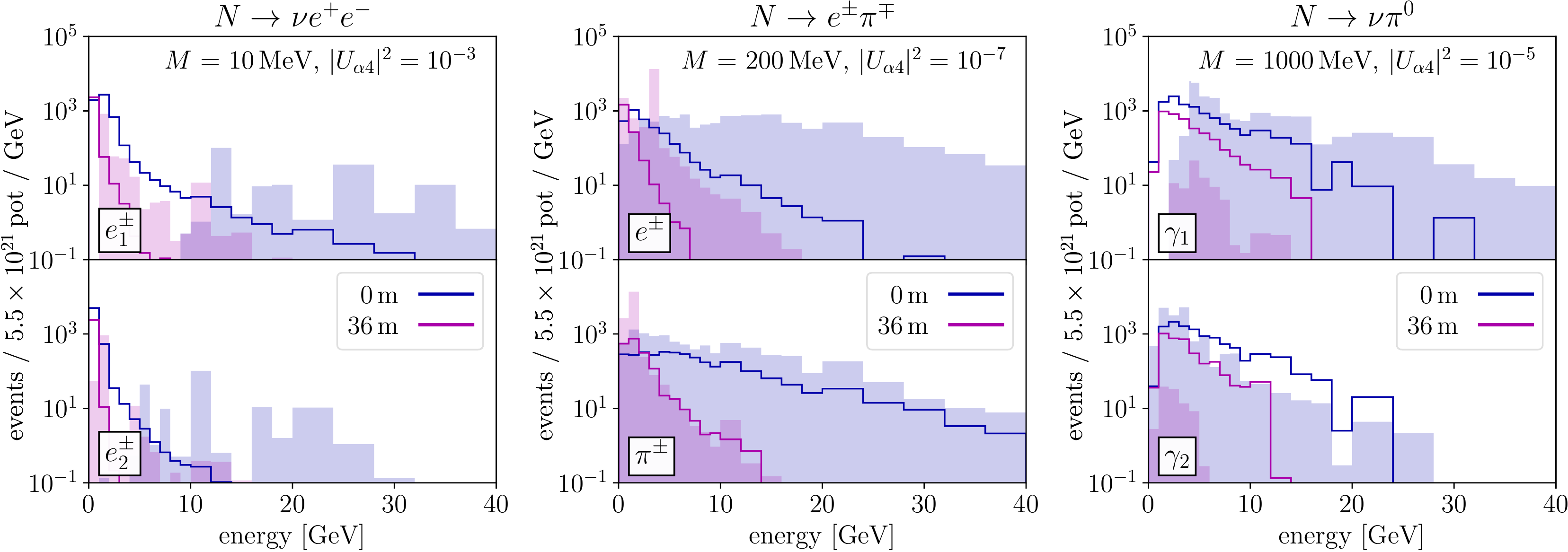}
  \caption{Signal and background event spectra at the DUNE near detector
    complex for specific HNL decay modes.  Blue and purple unshaded histograms
    correspond to the number of signal events per GeV at the on-axis location
    and at \SI{18}{m} off-axis, respectively. Shaded histograms represent the
    corresponding backgrounds. The three panels correspond to the most
    sensitive decay channels in the absence of backgrounds
    at $M = \SI{10}{MeV}$ (left panel), $M = \SI{200}{MeV}$
    (middle panel), and $M = \SI{1000}{MeV}$ (right panel), respectively, as indicated in
    the plots. For all decay channels included here ($N \to \nu e^+ e^-$ in the left
    panel, $N \to e^\pm \pi^\mp$ in the middle panel, $N \to \nu + (\pi^0 \to \gamma\gamma)$
    in the right panel), we show the spectra of the two visible final state
    particles separately. If the particles are identical (up to their charge,
    which we conservatively assume not to be measured), the upper histogram
    corresponds to the more energetic one, while the lower histogram is for the
    softer particle. Note that the sensitivity
    estimates discussed below are based on the full two-dimensional event
    distributions.  We observe that the signal-to-background ratio improves
    significantly when going off-axis.}
  \label{fig:hnl-spectra}
\end{figure}

In \cref{fig:hnl-spectra},
we compare the signal and background predictions for several HNL masses and
decay channels, both on-axis (blue) and off-axis (purple). We have chosen the
same representative HNL masses as in \cref{fig:hnl-flux}, and the decay channel
shown for each of them is the most sensitive one at this particular mass
(in the absence of backgrounds).
The values of the mixing matrix elements $|U_{\alpha 4}|^2$ (assumed to
be the same for
all flavors $\alpha = e,\;\mu,\;\tau$) are chosen
the 95\%~CL limits at the given masses.  All channels
shown here involve two visible final state particles, and we show the spectra
for both of them separately.  (In the case of $N \to \nu \pi^0$, the two
final state particles are the two photons from $\pi^0$ decay.)
If the two final state particles are identical (or identical up to
their charge, which we assume is not measured), the upper plot shows the spectrum
of the harder of the two, while the lower plot is
for the softer one.  We observe that, for all
channels, background levels at high energies are lowered by several orders of magnitude
when going off-axis. The signal spectra,
on the other hand, drop somewhat more slowly. This is true especially at low
energies, where the HNLs' forward boost
is relatively small and so their angular distribution is more isotropic
than the one of SM neutrinos, which are responsible for the background.

\subsection{Sensitivity to Heavy Neutral Leptons}
\label{sec:hnl-sensitivity}

We are now ready to estimate the sensitivity of the DUNE near detectors
to HNL decays.  We use again the log-likelihood function from \cref{eq:likelihood}, but
here we do so separately for ND-GAr and ND-LAr. This way, our analysis benefits from
the much better signal-to-background ratio in ND-GAr which comes from the fact that
the HNL signal scales with the detector volume, while the background scales with
its mass. The two independent $\chi^2$ values are only added up in the very end.
As the HNL decay final states that we consider here contain two visible particles,
we bin the events in two dimensions,
corresponding to the energies of the two particles.  This turns out to be very
important for optimizing the sensitivity. One problem with working with
two-dimensional histograms is that it is difficult to generate sufficient background
Monte Carlo statistics in all bins. This can be problematic because bins that contain
zero background events simply due to limited simulation statistics can lead to
sensitivity estimates that are too optimistic. To mitigate this problem,
we choose larger bins
at higher energies: our bin width is \SI{1}{GeV} for particle energies below \SI{10}{GeV},
\SI{2}{GeV} for particle energies up to \SI{20}{GeV}, and \SI{4}{GeV} for particle
energies up to \SI{40}{GeV}. To be on the safe side, we also identify bins in which
the background prediction is exactly zero. We then consider averages of the background
rate over increasingly larger neighborhoods of each such bin until we obtain a non-zero
rate. If that average rate is above 0.1~events, we exclude the problematic bin from our
analysis.

As systematic uncertainties, we include
a 10\% normalization error, which we assume to be uncorrelated between different HNL decay
channels and between the signal and background.  These uncertainties are described
in terms of nuisance parameters with Gaussian priors.  To obtain the final sensitivity,
our $\chi^2$ function is minimized over all nuisance parameters.

\begin{figure}
  \centering
  \includegraphics[width=\textwidth]{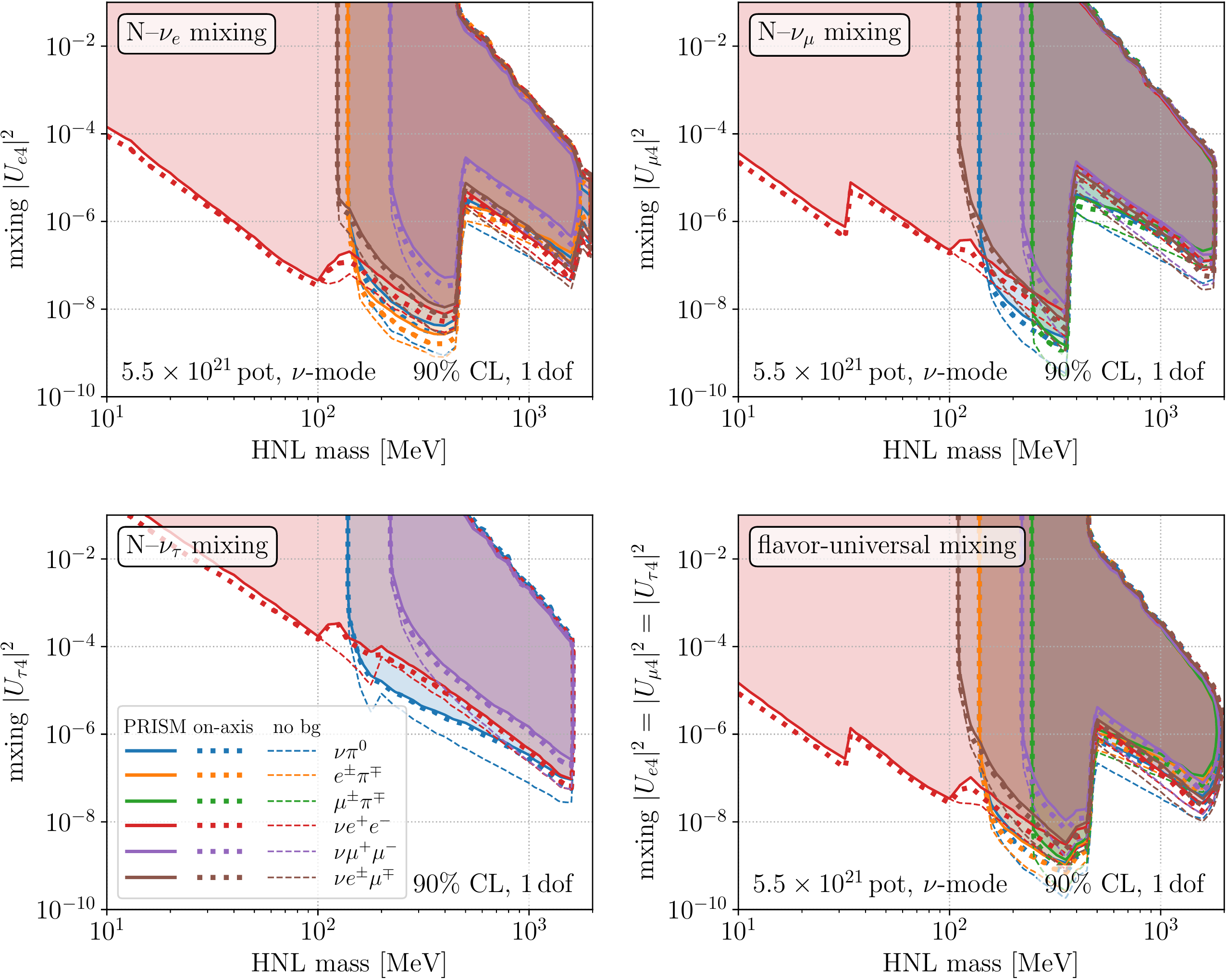}
  \caption{Sensitivity to HNLs in the DUNE near detectors
  ND-LAr and ND-GAr as function of HNL mass. Solid lines correspond to the
  DUNE-PRISM running strategy with data taking both on-axis and at six
  different off-axis locations (\SI{6}{m}, \SI{12}{m},
  \SI{18}{m}, \SI{24}{m}, \SI{30}{m}, \SI{36}{m}), with
  equal exposure for each location.  The total number of
  protons on target assumed here is $\num{5.5e21}$, corresponding to 5~years of
  running in neutrino mode.  Dotted lines show the sensitivity for on-axis data
  taking only, and thin dashed lines have been computed from a DUNE-PRISM
  analysis that neglects backgrounds.  The four panels correspond to different
  HNL coupling structures, namely mixing with $\nu_e$ only (top left panel), with
  $\nu_\mu$ only (top right panel), with $\nu_\tau$ only (bottom left panel), and
  flavor-universal mixing $|U_{e4}|^2 = |U_{\mu 4}|^2 = |U_{\tau 4}|^2 \equiv
  |U|^2$ (bottom right panel).
  }
  \label{fig:hnl-sensitivity-1}
\end{figure}

\begin{figure}
  \centering
  \vspace{-0.5cm}
  \includegraphics[width=0.6\textwidth]{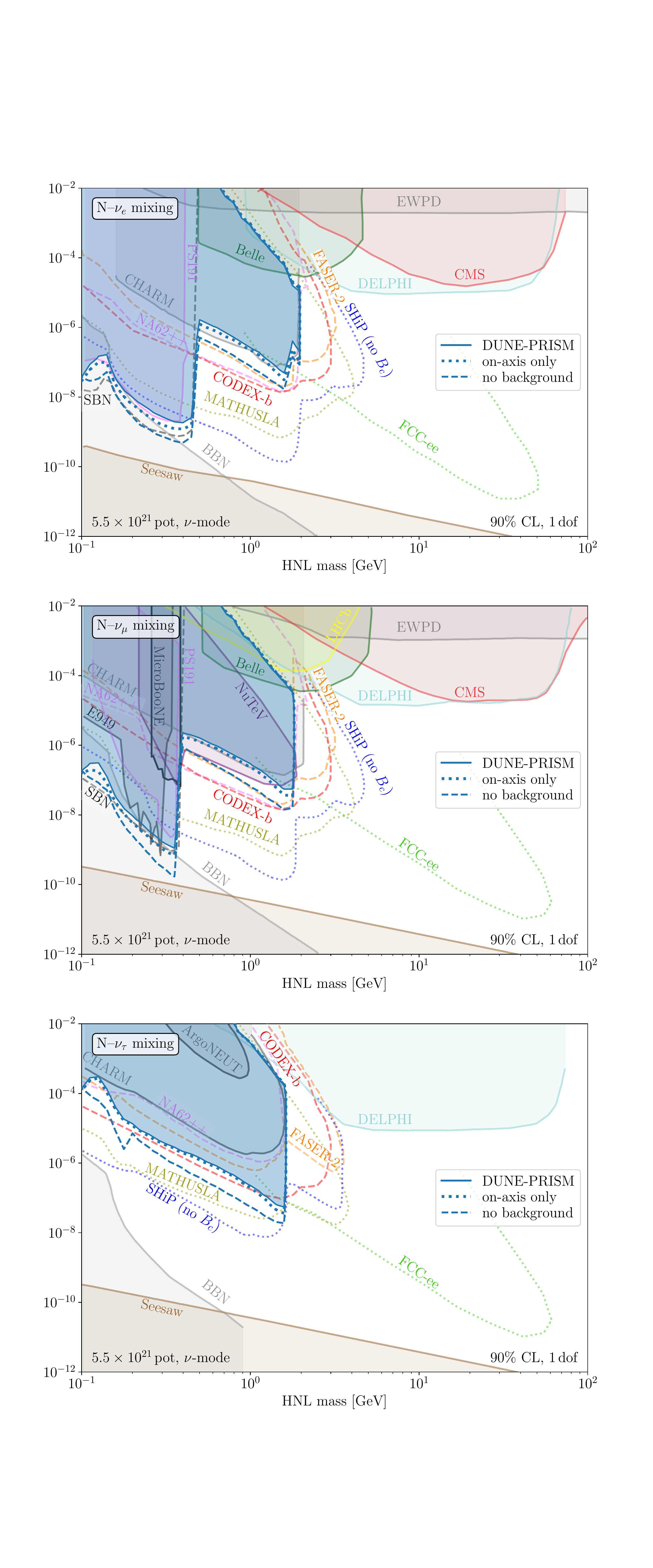}
  \caption{The DUNE near detectors compared to other experiments sensitive to
    HNLs.  The top, middle, and bottom panels show the
    sensitivity to the squared mixing matrix elements $|U_{e4}|^2$, $|U_{\mu
    4}|^2$, and $|U_{\tau 4}|^2$, respectively, assuming that only one of them
    is non-zero at a time.  In this plot, the decay channels $\nu e^+ e^-$,
    $\nu \mu^+ \mu^-$, $\nu e^\pm \mu^\mp$, $\nu \pi^0$, $\pi^\pm e^\mp$,
    $\pi^\pm \mu^\mp$ are combined. As in \cref{fig:hnl-sensitivity-1},
    solid blue curves show the sensitivity of DUNE-PRISM ($\SI{5.5e21}{pot}$
    in neutrino mode, equally split between the on-axis position and six
    different off-axis locations). Dotted blue contours show results for
    on-axis running only, and thin dashed blue curves represent a hypothetical
    background-free analysis.  The contours shown in the background correspond to existing limits (filled) and sensitivities of planned experiments (unfilled dashed and dotted).  These limits are taken from the compilation in ref.~\cite{Beacham:2019nyx} and from
    refs.~\cite{Ballett:2016opr,MicroBooNE:2019izn,ArgoNeuT:2021clc,Boiarska:2021yho}.}
  \label{fig:hnl-sensitivity-2}
\end{figure}

Our results are shown in \cref{fig:hnl-sensitivity-1,fig:hnl-sensitivity-2}.
The four panels in \cref{fig:hnl-sensitivity-1} correspond to different assumption
on the HNL couplings (coupling to $\nu_e$ only, to $\nu_\mu$ only, to
$\nu_\tau$ only, and to all three active neutrino flavors universally).
We observe rich structure in these plots: first, we see that the
projected exclusion regions have the wedge shape that is typical for
long-lived particle searches. At too small mixing, HNL production
is suppressed beyond the detectable level and HNLs are so long-lived that
they mostly decay far beyond the detector. At too large mixing, HNLs
are abundantly produced, but most of them decay before reaching
the detector.  Away from kinematic thresholds, the
upper limits on the mixing matrix elements $|U_{e4}|^2$,
$|U_{\mu 4}|^2$, $|U_{\tau 4}|^2$ (bottom edges of the wedge-shaped regions in
\cref{fig:hnl-sensitivity-1}) scale roughly as $|U_{\alpha 4}|^2_\text{limit}
\propto M^{-3}$. This can be understood as follows: the HNL production
rate scales as $|U_{\alpha 4}|^2 M^2$ over wide mass ranges, with the proportionality
to $M^2$ reflecting the chiral suppression that otherwise affects many
leptonic meson decays.  The HNL decay rate in the HNL rest frame scales
as $|U_{\alpha 4}|^2 M^5$, as can be seen from dimensional analysis. In the
laboratory frame, this scaling changes to $|U_{\alpha 4}|^2 M^6$ due to
relativistic time dilation. Simultaneously, the opening angle of the HNL
beam coming from its Lorentz boost grows with $M$, implying that the fraction
of HNLs crossing the detector drops as $M^{-2}$. Overall, these
arguments show that the experimental count rate scales as
$|U_{\alpha 4}|^4 M^6$, with deviations being observed close to
kinematic thresholds. Moreover, the geometric scaling factor
is not always exactly $M^{-2}$, depending on how exactly the flux in a
given channel drops with angle.

Let us now discuss the kinematic thresholds visible in \cref{fig:hnl-sensitivity-1}.
For instance, for $N$--$\nu_e$ coupling only
(top left panel of \cref{fig:hnl-sensitivity-1}), we see that, at
$M \sim m_\pi \sim \SI{140}{MeV}$, HNL production in pion decays becomes kinematically
forbidden, leading to a dent in the sensitivity in the $N \to \nu e^+ e^-$
channel.  Simultaneously, however,
new HNL decay modes ($N \to e^\pm \pi^\mp$ and $N \to \nu \pi^0$) become
allowed, compensating to some extent for this loss of sensitivity due to
their large branching ratios. The next kinematic threshold occurs at
$M \sim m_K \sim \SI{490}{MeV}$, when also HNL production in kaon decays
becomes forbidden. Beyond this threshold, only HNLs from charm decay
contribute, but since charm production in DUNE/LBNF is relative inefficient
due to the low primary proton energy of \SI{120}{GeV}, this leads to
a significant drop in sensitivity. Nevertheless, the DUNE near detectors will
be able to probe a large chunk of parameter space up to $M \sim m_{D_s}
\sim \SI{2}{GeV}$, where also HNL production in charm decays becomes forbidden.

For HNL couplings to $\nu_\mu$ (top right panel of \cref{fig:hnl-sensitivity-1}),
all thresholds are shifted by about $m_\mu \sim \SI{100}{GeV}$ because each
HNL needs to be produced together with a muon.  For HNL couplings exclusively
to $\nu_\tau$ (bottom left panel of \cref{fig:hnl-sensitivity-1}), production
thresholds do not play as important a role because charm decays are the
only production mode at all HNL masses due to the requirement of producing
a $\tau$ alongside the HNL. The small kink visible at $M \sim \SI{190}{MeV}$
can be understood from the $D_s$--$\tau$ mass difference. Beyond the kink,
HNL production in $D_s \to N + \tau$ decays is kinematically forbidden, leaving
only the off-shell decays $D_s \to \nu_\tau +
(\tau^* \to N e \nu_e,\,N \mu \nu_\mu,
N + \text{hadrons})$ as viable HNL production modes.

Comparing on-axis and off-axis sensitivities, we notice that going off-axis
does not lead to significant benefits, in spite of the much better signal-to-background
ratio.  The reason is that, with the cuts discussed above, background suppression
is fairly effective even on-axis. It is important to note, though, that
going off-axis also does not significantly harm the sensitivity for any channel,
in spite of the lower off-axis fluxes.
This means that
the search for HNLs can be carried out truly parasitically to DUNE's main
oscillation program. No matter where the various near detectors are placed at
any given moment, they will contribute in a useful way to the HNL sensitivity.
Notably, the SAND (System for On-Axis Neutrino Detection) beam monitor, which
will always remain on-axis and has not been included in our estimates, can be
harnessed for the HNL search as well.

In \cref{fig:hnl-sensitivity-2}, we compare DUNE's sensitivity to an array
of current and future HNL constraints, combining all six HNL decay channels
analyzed above. We do so for the same three scenarios
as in \cref{fig:hnl-sensitivity-1}: on-axis running only,
the realistic DUNE-PRISM running strategy with equal amounts
of data collected at seven
different locations relative to the beam axis (\SI{0}{m}, \SI{6}{m}, \SI{12}{m},
\SI{18}{m}, \SI{24}{m}, \SI{30}{m}, \SI{36}{m}~\cite{DUNEfluxes}),
and a hypothetical background-free DUNE-PRISM search.
The comparison reveals that DUNE will be able to somewhat improve
on existing limits for HNL couplings to $\nu_e$ and $\nu_\mu$, and will
go far beyond what is currently possible for HNL couplings to $\nu_\tau$.
(While this conclusion hinges to some extent on our treatment of charm
production in LBNF, we are confident that, given the very conservative
choice we have made for the charm production cross section, the sensitivity
of the real experiment will likely be better than our estimate.)
Note that the strongest existing limits on HNL mixing with $\nu_\tau$
shown in \cref{fig:hnl-sensitivity-2} are based on CHARM data \cite{CHARM:1985nku},
which has recently been reanalyzed~\cite{Boiarska:2021yho}.  We also
remind the reader that the limits from PS191 \cite{Bernardi:1987ek} shown here
have recently been called into question~\cite{Gorbunov:2021wua} and may be
weaker than indicated in \cref{fig:hnl-sensitivity-2}. This would make
future DUNE limits on HNL mixing with $\nu_\mu$ even more important.

Comparing to planned experiments that may happen on a timescale similar to
DUNE, we note that the ones that are most competitive with DUNE at low HNL
masses (below the kaon threshold) are the Fermilab short-baseline experiments
(``SBN''), that is MicroBooNE, SBND, and ICARUS.  Other planned experiments
(FASER-2, CODEX-b, NA62++, MATHUSLA, SHiP, FCC-ee) probe higher HNL masses than DUNE
because they operate at higher beam energies. They are, however, not able to
reach the luminosities achievable in DUNE.

We should also keep in mind that the limits shown here are conservative
in several ways, in particular they are based on only 5~years of data taking,
and we have assumed no charge identification capabilities.

\section{Summary and Conclusions}
\label{sec:summary}

In this work, we have explored the benefits of the DUNE-PRISM detectors
for exploring physics beyond the SM, namely two models of
light DM and a scenario featuring heavy neutral leptons
(or sterile neutrinos).  We have focused in particular
on the capability of these detectors to move in and out of the beam axis
and have found that exploiting this capability is highly beneficial in some
scenarios, and never disadvantageous.

For MeV--GeV-scale scalar DM coupled via a dark photon with kinetic mixing,
we have found that a detector placed off-axis will see a significant reduction
in backgrounds from SM neutrino scattering. While also the signal
due to DM--electron scattering is suppressed, the signal-to-background
ratio is improved significantly. This implies that a realistic DUNE-PRISM
running strategy that combines on-axis and off-axis measurements will be
ideal for constraining such a scenario. This is especially true if a simple
cut\,\&\,count analysis is performed; for a full spectral analysis,
an on-axis-only measurement would perform equally well as DUNE-PRISM. Overall,
a 5-year run of DUNE-PRISM (\SI{5.5e21}{pot}) will be able to improve existing
limits on dark photon-mediated light DM by up to a factor of a few,
as shown in \cref{fig:dp-limits}.

If light scalar DM couples through a leptophobic $Z'$ gauge boson
rather than a dark photon, we are in the interesting situation that, for
some parameter ranges (namely those with relatively heavy $Z'$), the
DM flux increases away from the beam axis, while the background
from SM neutral current neutrino interactions falls off.
Consequently, the DUNE-PRISM strategy is always advantageous compared
to an on-axis-only run, see \cref{fig:leptophobic-limits}. After 5~years,
DUNE-PRISM limits will be competitive with existing ones, and better in some
parameter regions.

Turning finally to  heavy neutral leptons (\cref{fig:hnl-sensitivity-2}),
we find once again that taking data both on-axis and off-axis as in DUNE-PRISM
never hurts the sensitivity. For some decay channels, especially those with
large backgrounds like $N \to \nu \pi^0$, off-axis running is highly
beneficial. For channels where backgrounds are lower or can be
well distinguished from the signal by using spectral information, the sensitivity
is similar off-axis and on-axis. Compared to existing limits, DUNE-PRISM
after 5~years will be competitive and in some parameter regions better than
existing limits for heavy neutral lepton mixing with $\nu_e$ and $\nu_\mu$.
Significant new territory will be covered for couplings to $\nu_\tau$ thanks
to charm production in the DUNE/LBNF target. 

We conclude that the DUNE-PRISM detectors are a versatile new tool for
probing numerous extensions of the SM. Their unique capability
of moving off-axis will only improve the sensitivity to such scenarios, implying
that a rich program of new physics searches can be carried out concurrently
with DUNE's neutrino oscillation program without requiring an adaptation of
the running strategy.

\section*{Acknowledgments}

We are deeply grateful to Tommaso Boschi for answering numerous questions about his {\sc NuShock} code \cite{NuShock}, to Laura Fields for useful discussions about the DUNE fluxes, and to Pedro Machado for helping us in the comparison of our results to the results of ref.~\cite{DeRomeri:2019kic}. We have benefited from useful discussions with Gaia Lanfranchi, Albert de Roeck, and the members of the DUNE-HNL working group.  This work has been partly funded by the German Research Foundation (DFG) in the framework of the PRISMA+ Cluster of Excellence and by the European Research Council (ERC) under the European Union's Horizon 2020 research and innovation programme (grant agreement No.\ \texttt{637506}, ``$\nu$Directions''). The research of L.\ B.\ is supported in part by the Swiss National Foundation under Contracts No.\ \texttt{200020\_188464} and No.\ \texttt{IZSAZ2\_173357}.

\appendix
\section{Simulation of Dark Matter Production in the Dark Photon Model}
\label{sec:DPcmp-deromeri}
    
In the context of the dark photon model introduced in \cref{sec:dark-photon}, we have compared our signal and background predictions with those of ref.~\cite{DeRomeri:2019kic}.  We have found that our estimate of the background due to neutrino--electron scattering agrees well with the results of ref.~\cite{DeRomeri:2019kic} in both the neutrino- and anti-neutrino mode. In the following, we focus on the estimate of the number of signal events. 
    
\begin{figure}
    \includegraphics[scale=0.9]{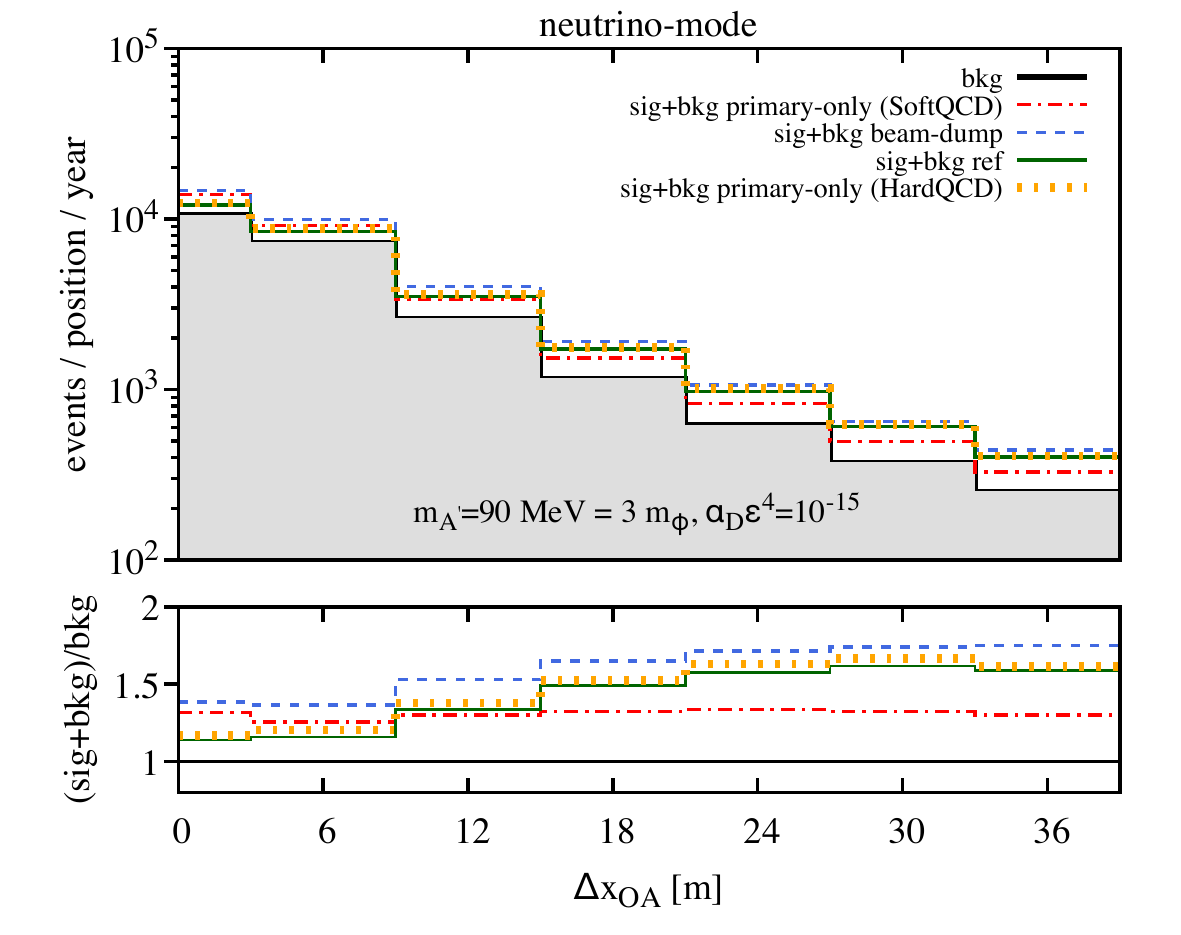}
    \caption{Same as \cref{fig:mesons-cmp}, but including the result of ref~\cite{DeRomeri:2019kic} in solid green, and our own \emph{primary-only (HardQCD)} sample in dotted orange.}
    \label{fig:mesons-cmp-ref}
\end{figure}

As discussed in sec.~\ref{sec:dp-production} a crucial aspect in the simulation of the signal events is the modeling of the light meson spectra from the DUNE/LBNF target. In ref.~\cite{DeRomeri:2019kic}, only the production of mesons in the primary proton interaction was considered, and {\sc Pythia} was used as the main simulation tool.  In \cref{fig:mesons-cmp-ref} we compare our results with those of fig.~3 of ref.~\cite{DeRomeri:2019kic} using the same benchmark point $\mAp = \SI{90}{MeV} = 3 m_\phi$, $\epsilon^4\alpha_D = 10^{-15}$. Contrary to our expectation, we observe that our \emph{primary-only (SoftQCD)} curve fails to accurately reproduce the results of that paper (solid green line) at any value of $\Delta x_\text{OA}$. As anticipated in \cref{sec:dp-production}, the source of the discrepancy is that the authors of ref.~\cite{DeRomeri:2019kic} have used a different configuration for their {\sc Pythia} simulation. More precisely, they activated the {\tt HardQCD:All} flag, and if we do the same (dotted orange curve in \cref{fig:mesons-cmp-ref}), we are indeed able to reproduce their results.

\begin{figure}
    \includegraphics[scale=0.9]{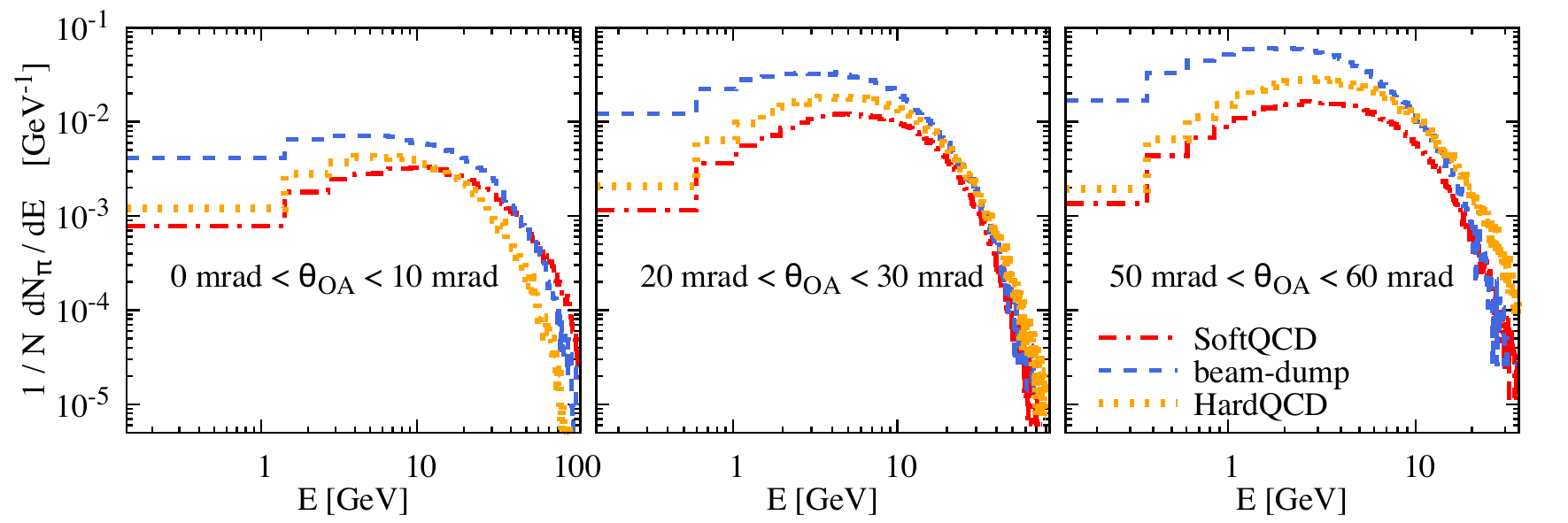} 
    \caption{Energy spectrum of $\pi^0$s produced in the beam dump in three angular windows: $0<\theta_\text{OA}<10\,$mrad (left panel), $20\,$mrad$<\theta_\text{OA}<30\,$mrad (middle panel) and $50\,$mrad$ <\theta_\text{OA}<60\,$mrad (right panel). The three curves correspond to the three meson samples introduced in the main text: \emph{primary-only (SoftQCD)} in red dot-dashed, \emph{beam-dump} in blue dashed \emph{primary-only (HardQCD)} in orange dotted.}
    \label{fig:mesons-spectra}
\end{figure}

We remark that using the {\tt SoftQCD:All} flag is preferable for modeling the proton--proton primary interactions for the case of relatively low energy fixed target experiments such as DUNE/LBNF. (This was shown in ref.~\cite{Berryman:2019dme}, which appeared after ref.~\cite{DeRomeri:2019kic}.) In fact, \emph{primary-only (HardQCD)} events tend to be much softer and have a larger angular spread than what is expected from primary proton--proton interactions, as also show in \cref{fig:mesons-spectra}. Hence, the \emph{primary-only (HardQCD)} sample underestimates the number of signal events on-axis, while being coincidentally similar to the \emph{beam-dump} sample when going off-axis.
    
\section{Statistical Analysis}
\label{sec:statistics}

In the following, we describe the methodology used to derive sensitivity limits from our predicted signal and background rates in the analyses of \cref{sec:dark-photon,sec:leptophobic,sec:hnl} in more detail. As stated in these sections we use standard frequentist techniques to test the
signal\,+\,background hypothesis against simulated background-only data. We employ a Poissonian log-likelihood function, i.e.\
\begin{multline}
  -2 \log\mathcal{L}(\vec\Theta, \vec{X})
     \equiv  2 \sum_{j=1}^{n_\text{pos}} \sum_{i=1}^{n_\text{bins}}  \bigg[
               B_{ij}(\vec{X}) + S_{ij}(\vec\Theta,\vec{X}) - B_{ij}(0)
             + B_{ij}(0) \log\bigg(
                        \frac{B_{ij}(0)}{B_{ij}(\vec{X}) + S_{ij}(\vec\Theta,\vec{X})}
                      \bigg) \bigg] \\
           + \sum_{c = S,B} \Bigg\{
             \bigg(\frac{X^\text{norm}_{c,\text{correl}}}{\sigma_\text{correl}}\bigg)^2
           + \bigg(\frac{X^\text{tilt}_{c,\text{correl}}}{\sigma_\text{correl}}\bigg)^2
           + \sum_{j=1}^{n_\text{pos}} \bigg[
               \bigg(\frac{X^\text{norm}_{c,j}}{\sigma_\text{pos}}\bigg)^2
             + \bigg(\frac{X^\text{tilt}_{c,j}}{\sigma_\text{pos}}\bigg)^2
             \bigg] \Bigg\} \,.
  \label{eq:likelihood}
\end{multline}
The signal and background rates in the $i$-th energy bin at the $j$-th off-axis
position including systematic biases, $S_{ij}(\vec\Theta,\vec{X})$ and
$B_{ij}(\vec{X})$, are defined in terms of the rates without biases,
$S_{ij}(\vec\Theta,0)$ and $B_{ij}(0)$, according to
\begin{align}
  \begin{split}
    S_{ij}(\vec\Theta, \vec{X}) &=
                       \big[1 + X^\text{norm}_{S,\text{correl}} \big]
		       \big[1 + X^\text{norm}_{S,j} \big]
		       \big[1 + X^\text{tilt}_{S,\text{correl}}  (-1,\dots,1)_i \big]
		       \big[1 + X^\text{tilt}_{S,j} (-1,\dots,1)_i \big]
                       S_{ij}(\vec\Theta,0) \,,
								    \\[0.2cm]
    B_{ij}(\vec{X}) &= \big[1 + X^\text{norm}_{B,\text{correl}} \big]
		       \big[1 + X^\text{norm}_{B,j} \big]
		       \big[1 + X^\text{tilt}_{B,\text{correl}}  (-1,\dots,1)_i \big]
		       \big[1 + X^\text{tilt}_{B,j} (-1,\dots,1)_i \big]
                       B_{ij}(0) \,.
  \end{split}
\end{align}
We collectively denote
the vector of physical model parameters $\vec\Theta$ and the vector of nuisance
parameters $\vec{X}$. The nuisance parameters $X^\text{norm}_{S,\text{corr}}$ and $X^\text{norm}_{S,j}$
describe systematic normalization uncertainties in the signal, while the parameters
$X^\text{tilt}_{S,\text{correl}}$ and $X^\text{tilt}_{S,j}$ parameterize spectral
``tilts'': their effect is to pivot the spectrum about its midpoint.
The meaning of the corresponding parameters affecting the background
 ($X^\text{norm}_{B,\text{corr}}$, $X^\text{norm}_{B,j}$,
$X^\text{tilt}_{B,\text{correl}}$ and $X^\text{tilt}_{B,j}$) is analogous.
(Note that tilt errors are not considered in the HNL analysis of \cref{sec:hnl}.)
All systematic errors are treated as Gaussian and are constrained by the pull terms
in the second row of \cref{eq:likelihood}. As systematic normalization and tilt
uncertainties we assume $\sigma_\text{pos} = 1\%$ (uncorrelated between
different on-/off-axis positions) and $\sigma_\text{correl} = 10\%$
(correlated among all positions).
The sums in the first line of \cref{eq:likelihood} run over $n_\text{bins}=80$ ($n_\text{bins}=57$) energy bins, equally spaced in the interval $[0,20]\,\si{GeV}$ ($[3,60]\,\si{GeV}$) in case of the dark photon model (leptophobic model), and over $n_\text{pos} = 7$ on- and off-axis
positions (\SI{0}{m}, \SI{6}{m}/\SI{10.45}{mrad}, \SI{12}{m}/\SI{20.90}{mrad},
\SI{18}{m}/\SI{31.36}{mrad}, \SI{24}{m}/\SI{41.81}{mrad},
\SI{30}{m}/\SI{52.26}{mrad}, \SI{36}{m}/\SI{62.72}{mrad}). For comparison we also present results for an on-axis-only run, or for data taken at a fixed
off-axis location. In this case, of course, we set $n_\text{pos} = 1$.
For HNL searches, we use the two-dimensional binning described in \cref{sec:hnl-sensitivity}.

The statistical significance at which a given parameter point $\vec\Theta$
is excluded can be estimated from the log-likelihood ratio \cite{Cowan:2010js}
\begin{align}
  Z(\vec\Theta) \equiv -2 \, {\log}{\bigg(
                       \frac{\mathcal L(\vec\Theta, \doublehat{\vec X})}
                            {\mathcal L(\hat{\vec\Theta}, \hat{\vec X})} \bigg)} \,,
  \label{eq:Z-Theta}
\end{align}
where $(\hat{\vec\Theta}, \hat{\vec X})$ is the combination of model parameters and
nuisance parameters that maximizes the likelihood (minimizes $\chi^2$),
and $\doublehat{\vec X}$ are
the nuisance parameters that maximize the likelihood for fixed $\vec\Theta$.
$Z(\vec\Theta)$ follows a $\chi^2$ distribution, with the number of degrees of
freedom equal to the dimension of $\vec\Theta$. Consequently, the $n\sigma$
exclusion region can be estimated by comparing $Z(\vec\Theta)$ to the
$n\sigma$ quantile of that $\chi^2$ distribution.

However, experimental collaboration often present their results as
constraints on the ``signal strength'' $\mu$, which is defined as an overall,
energy-independent rescaling factor of the event rate relative to some
reference point.  In the dark photon model, the signal strength can be taken as
the product of coupling constants $\mu \equiv \epsilon^4 \alpha_D$, while in the leptophobic model it is $\mu \equiv g_Z^6$. In this
way of analyzing the data, $Z(\vec\Theta)$ is replaced by a one-parameter
function
\begin{align}
  Z(\mu) \equiv -2 \, {\log}{\bigg(
                       \frac{\mathcal L(\mu, \doublehat{\vec X})}
                            {\mathcal L(\hat{\mu}, \hat{\vec X})} \bigg)} \,,
  \label{eq:Z-mu}
\end{align}
whose values are then compared to the $\chi^2$ distribution
with one degree of freedom. In \cref{eq:Z-mu}, $(\hat{\mu}, \hat{\vec X})$ is once again the parameter point in $(\mu, \vec{X})$ at which
the likelihood is maximal.  In a signal strength analysis, however, only
$\mu$ is varied and all other model parameters are kept fixed.
The statistical question answered by a signal
strength analysis is ``What is the constraint on the signal strength, assuming
we already know all other model parameters.'' A multi-parameter fit, on the
other hand, asks ``What are the preferred parameter regions, assuming the
chosen model is the correct one, but we do not have any prior information
on any of its parameters.'' To make our results comparable to those
shown in the literature, we use signal strength analyses throughout.


\bibliography{refs}
\bibliographystyle{JHEP}

\end{document}